\documentclass[a4paper,11pt]{article}
\usepackage{jheppub}
\usepackage{amsmath,amssymb,bm,graphicx,bbold,epsf,colordvi}
\usepackage{lipsum}
\usepackage{soul}
\usepackage[normalem]{ulem}
\usepackage{braket}
\usepackage{bm}
\allowdisplaybreaks 
\usepackage{graphicx}
\usepackage{subcaption}
\usepackage[dvipsnames]{xcolor}
\makeatletter
\def\@fpheader{~}
\makeatother

\usepackage{comment}

\usepackage{color}

\newcommand{\nn}{\notag}
\newcommand{\df}{\mathrm{d}}
\newcommand{\tr}{\operatorname{tr}}

\newcommand{\sech}{\operatorname{sech}}
\newcommand{\mO}{\mathcal{O}}

\DeclareMathOperator*{\SumInt}{
\mathchoice
  {\ooalign{$\displaystyle\sum$\cr\hidewidth$\displaystyle\int$\hidewidth\cr}}
  {\ooalign{\raisebox{.14\height}{\scalebox{.7}{$\textstyle\sum$}}\cr\hidewidth$\textstyle\int$\hidewidth\cr}}
  {\ooalign{\raisebox{.2\height}{\scalebox{.6}{$\scriptstyle\sum$}}\cr$\scriptstyle\int$\cr}}
  {\ooalign{\raisebox{.2\height}{\scalebox{.6}{$\scriptstyle\sum$}}\cr$\scriptstyle\int$\cr}}
}

\title{\boldmath Resummed azimuthal decorrelation and transverse momentum imbalance of dijets at the LHC}
\author[a]{Rong-Jun Fu}
\author[b]{, Rudi Rahn}
\author[a,c,d]{, Ding Yu Shao}
\author[e,f]{, Wouter J.~Waalewijn}
\author[g]{and Bin Wu}
\affiliation[a]{Department of Physics, Center for Field Theory and Particle Physics and Key Laboratory of Nuclear Physics and Ion-beam Application (MOE), Fudan University, Shanghai, 200433, China}
\affiliation[b]{University of Vienna, Faculty of Physics,\\Boltzmanngasse 5, A-1090 Wien, Austria}
\affiliation[c]{Shanghai Research Center for Theoretical Nuclear Physics, NSFC and Fudan University, Shanghai 200438, China}
\affiliation[d]{Center for High Energy Physics, Peking University, Beijing 100871, China}
\affiliation[e]{Nikhef, Theory Group,\\Science Park 105, 1098 XG, Amsterdam, The Netherlands}
\affiliation[f]{Institute for Theoretical Physics Amsterdam and Delta Institute for Theoretical Physics, University of Amsterdam,\\Science Park 904, 1098 XH Amsterdam, The Netherlands}
\affiliation[g]{Instituto Galego de F\'isica de Altas Enerx\'ias IGFAE, Universidade de Santiago de Compostela,\\E-15782 Galicia-Spain}

\emailAdd{rjfu23@m.fudan.edu.cn}
\emailAdd{rudi.rahn@univie.ac.at}
\emailAdd{dyshao@fudan.edu.cn}
\emailAdd{w.j.waalewijn@uva.nl}
\emailAdd{b.wu@cern.ch}

\abstract{
We present a theoretical study of the azimuthal decorrelation $\delta\phi$ and transverse momentum imbalance $q_T$ in dijet production at the LHC, offering intriguing insights into the dynamics of quantum chromodynamics. We define the jet axes using the recoil-free winner-take-all (WTA) recombination scheme. For the azimuthal decorrelation $\delta\phi$, this axis choice eliminates non-global logarithms (NGLs) entirely. For the transverse momentum imbalance $q_T$, NGLs emerge specifically in the small jet radius limit ($R \ll 1$). In this regime, the WTA scheme simplifies the theoretical framework by restricting jet radius logarithms to the soft sector. We derive factorization formulae for both observables within soft-collinear effective theory. To address the small-$R$ NGLs in the $q_T$ distribution, we refactorize the soft function into global soft, collinear-soft, and ultra-collinear-soft modes. We perform the resummation of global large logarithms $\ln(\delta\phi)$ and $\ln(q_T/Q)$ up to next-to-next-to-leading logarithmic accuracy. For the $q_T$ distribution, this is combined with a leading-logarithmic resummation of the non-global $\ln R$ terms. We match our predictions to leading fixed-order $\mathcal{O}(\alpha_s^3)$ calculations. We also numerically investigate the structure of the first subleading power corrections. Comparisons with \texttt{PYTHIA}~8 simulations demonstrate that the observables we consider are robust against non-perturbative multi-parton interactions and hadronization effects.
}

\begin{document}
\preprint{UWThPh 2026-3}
\maketitle
\flushbottom

\section{Introduction}
\label{sec:intro}

The study of jet production at hadron colliders provides a fundamental test of quantum chromodynamics (QCD) and serves as a crucial tool for probing the structure of the proton. Among the simplest yet most insightful observables in this context are those characterizing the correlations between high transverse momentum objects produced in a back-to-back configuration. For processes such as vector boson plus jet ($V$+jet) production or inclusive (i.e.,~without a jet veto) dijet production, momentum conservation in the transverse plane enforces this back-to-back alignment at the Born level. However, soft and collinear radiation induces deviations from this topology, generating imbalances in the transverse momentum $q_T$ or azimuthal angle $\delta\phi$. The azimuthal decorrelation is sensitive to the dynamics of QCD radiation and is of significant experimental interest at the LHC~\cite{Chatrchyan:2013tna, ATLAS:2016jxf, Sirunyan:2017jic, Aaboud:2017kff, CMS:2017ehl, ATLAS:2011kzm, CMS:2011hzb, CMS:2016adr, CMS:2017cfb, ATLAS:2018sjf, CMS:2019joc}. The transverse momentum imbalance is used, for example, in jet calibration~\cite{ATLAS:2012cse, CMS:2011shu}. Furthermore, single-spin asymmetries in back-to-back dijet production at polarized proton-proton colliders serve as crucial probes for the three-dimensional tomographic imaging of the proton and the study of quark and gluon transverse-momentum-dependent (TMD) parton distribution functions (PDFs)~\cite{Boer:2003tx, Bomhof:2007su, Qiu:2007ey, Vogelsang:2007jk, Kang:2020xez, Liu:2020jjv, STAR:2023xvk}. The consistent extraction of these TMD distributions is complicated by potential factorization-breaking effects~\cite{Collins:2007nk, Rogers:2010dm, Gaunt:2014ska, Zeng:2015iba, Catani:2011st, Forshaw:2012bi, Schwartz:2017nmr, Cieri:2024ytf, Henn:2024qjq, Guan:2024hlf}.

Precise theoretical predictions for these observables require the all-order resummation of large logarithms that emerge in the back-to-back limit. However, achieving high logarithmic accuracy for jet observables is notoriously difficult due to the presence of non-global logarithms (NGLs)~\cite{Dasgupta:2001sh}. These arise from soft radiation that is sensitive to the jet boundary, leading to complex factorization structures~\cite{Hatta:2013iba, Larkoski:2015zka, Caron-Huot:2015bja, Becher:2015hka, Becher:2016mmh}. Consequently, theoretical predictions for these TMD-like observables using standard jet definitions have traditionally been restricted to next-to-leading logarithmic (NLL) accuracy~\cite{Sun:2014gfa, Sun:2015doa, Liu:2018trl, Chien:2019gyf, Gao:2023ulg}, though recent theoretical efforts have begun extending this frontier toward next-to-next-to-leading logarithmic (NNLL)  precision~\cite{Balsiger:2019tne, Banfi:2021owj, Becher:2023vrh}.

In this paper, we study the azimuthal decorrelation $\delta\phi$ and the transverse momentum imbalance $q_T$ in inclusive dijet production. To address the challenge of NGLs, we adopt the winner-take-all (WTA) recombination scheme~\cite{Salam:WTAUnpublished, Bertolini:2013iqa} to define the jet axes. The WTA scheme is \emph{recoil-free}, meaning the direction of the jet axis is insensitive to soft recoil~\cite{Banfi:2008qs, Larkoski:2014uqa}. This property eliminates NGLs from the factorization of the azimuthal decorrelation $\delta\phi$, enabling high-precision resummation, as demonstrated in previous studies of $V$+jet production~\cite{Chien:2020hzh, Chien:2022wiq}.

The theoretical framework for $q_T$ is more nuanced. While the WTA scheme removes NGLs from the azimuthal decorrelation, the transverse momentum imbalance retains a sensitivity to NGLs in the small jet radius limit ($R \ll 1$), even under WTA definitions~\cite{Chien:2022wiq}. Building on recent developments in transverse momentum slicing for WTA jets~\cite{Fu:2024fgj}, we here investigate the logarithmic structure of the $q_T$ distribution. We show that the WTA scheme allows us to isolate the origin of these NGLs as jet radius logarithms ($\ln R$) within the soft sector. This enables a joint resummation framework within soft-collinear effective theory (SCET)~\cite{Bauer:2000ew, Bauer:2000yr, Bauer:2001ct, Bauer:2001yt, Bauer:2002nz, Beneke:2002ph}, where global logarithms $\ln(q_T/p_T)$ are resummed to NNLL accuracy, while the NGLs arising from the interplay between collinear-soft and ultra-collinear-soft modes are resummed to LL accuracy.

Here we extend the WTA-based resummation framework from $V$+jet to inclusive dijet production, which involves several non-trivial theoretical advancements: Unlike $V$+jet processes, the dijet sector is characterized by a large number of partonic channels and non-trivial color structures. This complexity necessitates a more sophisticated treatment of the hard function to describe the scattering process and the soft function to encode wide-angle radiation from both the incoming and outgoing hard partons. Furthermore, the role of linearly polarized gluons differs significantly. While they contribute at next-to-leading order (NLO) for massive vector boson plus jet production~\cite{Chien:2020hzh}, their impact in the dijet case only emerges at next-to-next-to-leading order (NNLO)~\cite{Fu:2024fgj}.

The resummation structure itself is also modified. For the two-component transverse momentum, the collinear and soft radiation confined within the jet contribute only via the momentum component perpendicular to the beam-jet plane. Additionally, the distinct treatment of soft radiation across the jet boundary introduces NGLs. In the WTA scheme, these are specifically identified as NGLs of the jet radius $R$. To resum them, we perform a refactorization of the soft function into global soft, collinear-soft, and ultra-collinear-soft modes, and we find their numerical impact to be relatively mild. Finally, we refine the fixed-order matching procedure. While $V$+jet matching is often complicated by large corrections from similar final states in electroweak sectors~\cite{Rubin:2010xp}, the dijet matching presented here provides a robust baseline for high-precision LHC phenomenology.

Our main results include a comprehensive phenomenological study at the LHC. We present NNLL resummed predictions for both $\delta\phi$ and $q_T$, matched to NLO fixed-order results generated by \texttt{NLOJET++}~\cite{Nagy:2001xb, Nagy:2001fj, Nagy:2003tz} and analyzed using \texttt{FASTJET}~\cite{Cacciari:2011ma}. We discuss the substantial reduction in theoretical uncertainties achieved by the resummation and analyze the impact of non-perturbative effects by comparing our predictions with \texttt{PYTHIA}~8~\cite{Bierlich:2022pfr} simulations. We find that both $\delta\phi$ and $q_T$ are robust against hadronization and multi-parton interactions.

The paper is organized as follows. Section~\ref{sec:kinematics} introduces the kinematic setup and the definitions of the observables. In section~\ref{sec:fact}, we derive the factorization formulae for $\delta\phi$ and $q_T$, with a detailed discussion of the small-$R$ refactorization for the $q_T$ soft function. The resummation of large logarithms and the evolution equations needed to accomplish this are presented in section~\ref{sec:resum}. In section~\ref{sec:numerical}, we present our numerical results, including uncertainty estimates, matching to fixed order, and comparisons with Monte Carlo simulations. We conclude in section~\ref{sec:conclusions}.

\section{Kinematics} \label{sec:kinematics}

We consider the inclusive dijet production process at hadron colliders,
$$
p(P_a) + p(P_b) \to J(p_1) + J(p_2) + X.
$$
Here, $p(P_{a,b})$ denote the incoming protons, $J(p_{1,2})$ are the two leading final-state jets, and $X$ represents unobserved hadronic radiation. At the Born level, this corresponds to a $2 \to 2$ partonic scattering process. Treating all partons as massless, we parameterize their four-momenta in the laboratory frame as 
\begin{align}
 &p_a^\mu = x_a P_a^\mu = x_a\frac{\sqrt{s}}{2}(1,0,0,1), \quad p_b^\mu = x_b P_b^\mu = x_b\frac{\sqrt{s}}{2}(1,0,0,-1),\quad \\
&p_1^\mu = p_{T}(\cosh{\eta_1},1,0,\sinh{\eta_1}), \quad p_2^\mu = p_{T}(\cosh{\eta_2},-1,0,\sinh{\eta_2}), 
\end{align}
where $s\equiv(P_a + P_b)^2$, and $p_T$ and $\eta_{1,2}$ are the transverse momentum and rapidities of the outgoing partons, which are back-to-back in the transverse plane. The partonic momentum fractions $x_{a,b}$ are fixed by 
\begin{equation}
x_a = \frac{p_T}{\sqrt{s}} (e^{\eta_1}+e^{\eta_2}),\quad
x_b = \frac{p_T}{\sqrt{s}} (e^{-\eta_1}+e^{-\eta_2}).
\end{equation}
The partonic Mandelstam variables are then given by
\begin{align}
\hat{s} &= (p_a+p_b)^2 = x_a x_b s = 2p_T^2[1+\cosh{(\eta_1-\eta_2)}], \label{eq:s_hat} \\
\hat{t} &= (p_a-p_1)^2 = -p_T^2 (1+e^{\eta_2-\eta_1}), \\
\hat{u} &= (p_a-p_2)^2 = -p_T^2 (1+e^{\eta_1-\eta_2}).
\end{align}

Beyond the Born approximation, jets are clustered using the anti-$k_T$ algorithm with the standard $E$-scheme. This work focuses on the back-to-back kinematic regime, where the total dijet transverse momentum $q_T \to 0$. In this limit, the cross section is sensitive to large logarithms, such as $\ln(q_T/Q)$ where $Q \sim p_T$, which spoil the convergence of fixed-order perturbation theory and require resummation.

A known complication of all-order resummation in such scenarios is the presence of NGLs. To address this issue, we define our observables using the winner-take-all (WTA) recombination scheme, rather than the standard $E$-scheme. The WTA scheme is recoil-free and is known to eliminate NGLs from the factorization for $\delta\phi$. For the  $q_T$ distribution, the scheme simplifies the framework as the remaining NGLs only depend on the jet radius $R$\footnote{The corresponding non-global structure for the Standard Jet Axis would involve a combination of NGLs of both jet radius and observable.}.
Explicitly, we measure the dijet transverse momentum imbalance $\vec{q}_T$ and the azimuthal acoplanarity $\delta\phi$ defined by
\begin{equation}
    \vec{q}_T \equiv \vec{p}_{1,\,T}^{\,\text{WTA}} + \vec{p}_{2,\,T}^{\,\text{WTA}},\quad \delta\phi \equiv |\pi-|\phi_1^{\text{WTA}}-\phi_2^{\text{WTA}}||.
\end{equation}
Here, $\vec{p}_{i,\,T}^{\,\text{WTA}}$ and $\phi_i^{\text{WTA}}$ are the transverse momentum and azimuthal angle of the jet axis defined by the WTA scheme. It is important to note that these kinematic quantities are distinct from the transverse momentum $\vec{p}_{i,\,T}$ and azimuthal angle $\phi_i$ that would be obtained using the standard $E$-scheme. While the change in transverse momentum and rapidity of the individual jets due to the scheme choice is rather small, the effect on $\vec{q}_T$ and $\delta\phi$ is not small because these observables involve large cancellations between the two almost back-to-back jets.

Even with this simplification, the resummations for $q_T$ and $\delta\phi$ remain distinct. As demonstrated in ref.~\cite{Fu:2024fgj}, the $q_T$ and $\delta\phi$ distributions possess different factorization structures. In section~\ref{sec:fact}, we will derive the factorization and resummation formulae for both of these observables.

\subsection{Kinematic analysis} \label{sec:kinematic_analysis}
According to the WTA-$p_T$ scheme defined in the transverse plane perpendicular to the beam axis, we can rewrite the observable $\vec{q}_T$ in a more explicit form as
\begin{align}
    \vec{q}_T &\equiv \vec{p}_{1,\,T}^{\,\text{WTA}} + \vec{p}_{2,\,T}^{\,\text{WTA}} \nn \\
    &=  \Biggl(\sum_{k\in \text{jet-}1} p_{k,\,T}\Biggr) \vec{n}_{W_1,\,T} + \Biggl(\sum_{k\in \text{jet-}2} p_{k,\,T}\Biggr)\, \vec{n}_{W_2,\,T} \,,
\end{align}
where $\vec{n}_{W_i,\,T}$ denotes the direction of the WTA axis of jet-$i$ ($i=1,2$) in the transverse plane. Here, we use $\sum_{k\in \text{jet}} p_{k,\,T}$ to represent the scalar sum of the transverse momentum of all the collinear partons inside the jet. The soft radiation inside the jet is power-suppressed compared with the collinear partons. Utilizing momentum conservation
\begin{align}
    \sum_{k\in \text{jet-}1} \vec{p}_{k,\,T} + \sum_{k\in \text{jet-}2} \vec{p}_{k,\,T} + \vec{p}_{a,\,T} + \vec{p}_{b,\,T} + \vec{p}_{s,\,T} = 0,
\end{align}
we obtain 
\begin{align} \label{eq:qTrewrite}
    \vec{q}_T =& \sum_{k\in \text{jet-}1} \left(p_{k,\,T}\, \vec{n}_{W_1,\,T} - \vec{p}_{k,\,T} \right) + \sum_{k\in \text{jet-}2}\left(p_{k,\,T}\, \vec{n}_{W_2,\,T} - \vec{p}_{k,\,T}\right) - \vec{p}_{a,\,T} - \vec{p}_{b,\,T} - \vec{p}_{s,\,T}  \nn\\
    =& -\sum_{k\in \text{jet-}1} \vec{p}_{k,\,\perp}^{\,(1)} -\sum_{k\in \text{jet-}2} \vec{p}_{k,\,\perp}^{\,(2)} - \vec{p}_{a,\,T} - \vec{p}_{b,\,T} - \vec{p}_{s,\,T},
\end{align}
where the reconstructed collinear transverse momentum $\vec{p}_{k,\,\perp}^{\,(i)}$ ($i=1,2$) is defined as
\begin{align}
    \vec{p}_{k,\,\perp}^{\,(i)} \equiv \vec{p}_{k,\,T}~ - ~p_{k,\,T}~\vec{n}_{W_i,\,T}, 
    \quad k\in \text{jet-}i.
\end{align}
Since $\vec{p}_{k,\,T}$ is collinear to the jet axis, one has 
\begin{equation} \label{eq:inner_prod}
    \vec{p}_{k,\,\perp}^{\,(i)} \cdot \vec{n}_{W_i,\,T} = 0 + \mathcal{O}\biggl(p_{k,\,T}\, (\phi_k - \phi_i^\text{WTA})^2\biggr)
\end{equation}
for collinear emissions in jet-$i$, with $\phi_k$ the emission's azimuthal orientation. Eq.~\eqref{eq:inner_prod} implies that at leading power $\vec{p}_{k,\,\perp}^{\,(i)}$ can be treated as perpendicular to the jet and the beam axes, which motivates the subscript $\perp$. The above derivation is true for $pp\to N$~jets. Furthermore, when we consider the $pp \to$ dijet process in a reference frame where the scattering plane lies in the $zx$-plane at Born level, we find that the transverse momentum of collinear radiation inside the jet is dominated by its $y$-component. In contrast, collinear radiation along the beam directions and soft radiation exhibit a full two-dimensional transverse-momentum dependence, as shown in eq.~\eqref{eq:qTrewrite}. 

Furthermore, if we focus from the outset on the $q_y$ observable (or equivalently, $\delta\phi$)\footnote{Note that in refs.~\cite{Chien:2022wiq, Gao:2023ulg, Fu:2024fgj} we used a different convention, using $q_x$ instead of $q_y$ for the component corresponding to $\delta \phi$.}, we can directly formulate the factorization expressions in the one-dimensional Fourier-conjugate space associated with $b_y$, as shown in eq.~\eqref{eq:fac_dphi}.

\section{Factorization and anomalous dimensions} \label{sec:fact}
In this section, we present the TMD factorization structures for the distributions of the azimuthal decorrelation $\delta\phi$ and the transverse momentum imbalance $q_T$, respectively. We also summarize the anomalous dimensions required for the resummation.

In this paper, we employ the light-cone decomposition with respect to the directions $n_i = (1,\vec{n}_i)$ and $\bar{n}_i = (1,-\vec{n}_i)$, where $|\vec{n}_i|=1$. An arbitrary momentum $p^\mu$ is decomposed as
\begin{align} \label{eq:lightcone_comp}
    p^\mu = (n_i\cdot p) \frac{\bar{n}_i^\mu}{2} + (\bar{n}_i\cdot p) \frac{n_i^\mu}{2} + p_\perp^\mu
        \equiv (n_i\cdot p, ~\bar{n}_i\cdot p, ~p_\perp^\mu)_{n_i \bar{n}_i}.
\end{align}
We further define the light-cone components using $\pm$ superscripts as $p^+ \equiv n_i\cdot p$ and  $p^- \equiv \bar{n}_i\cdot p$.

\subsection{Azimuthal decorrelation} \label{sec:dphi_fact}

\begin{figure}[tp]
\centering
\includegraphics[width=0.5\textwidth]{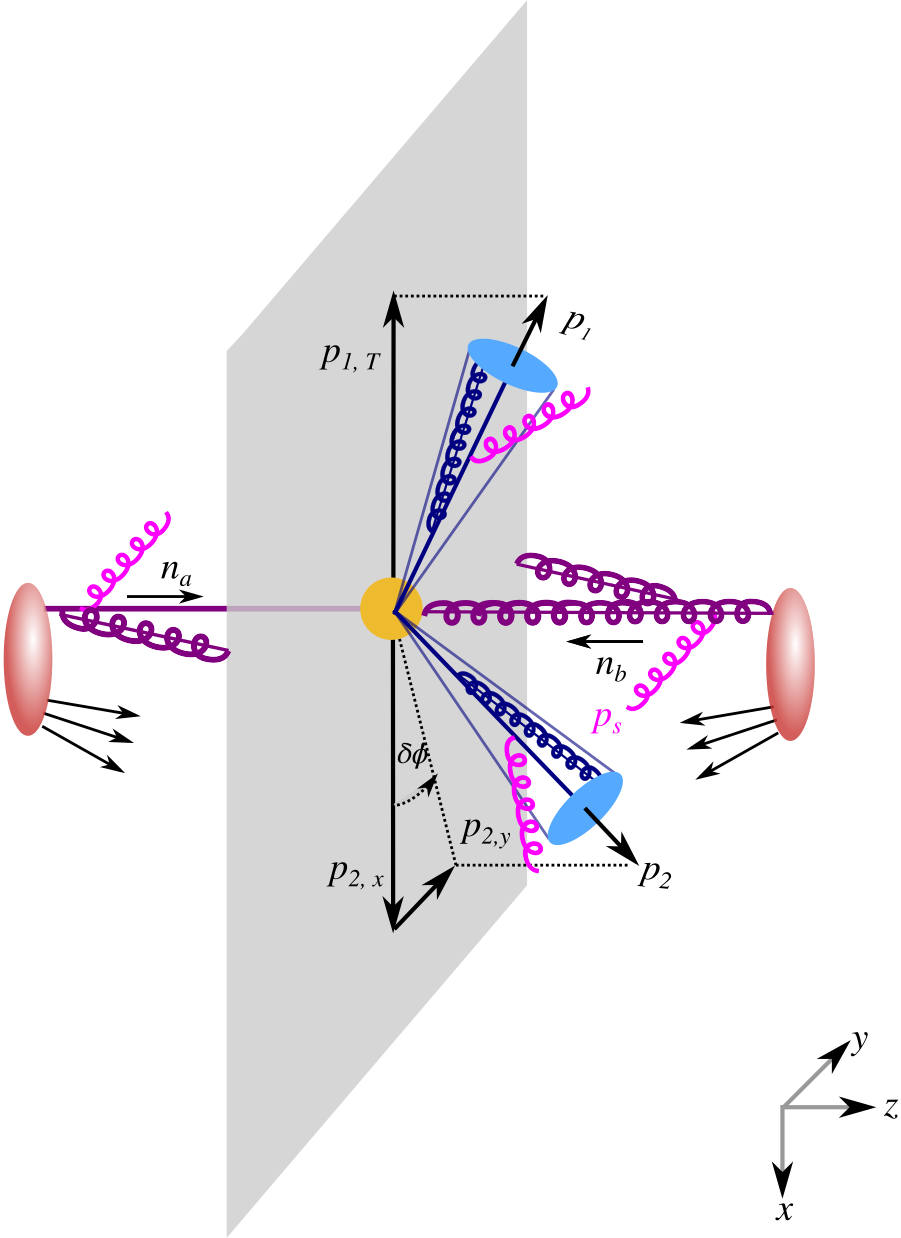}
\caption{Kinematics of the $pp \to $ dijet process in the transverse plane. For simplicity, the $x$-axis is aligned with the first jet, such that $\delta \phi$ is related to $q_y= p_{2,y}$. 
For the factorization theorem, it is more natural to consider a reference frame where the $x$-axis lies along the direction of the two back-to-back partons that initiate the jets. In this frame, the jet functions encode the offset between these parton directions and the jet axes, while the soft and beam functions capture the recoil induced by soft and initial-state collinear radiation. This choice of frame does not alter the factorization structure.}
\label{fig:coordinates}
\end{figure}

In the back-to-back limit ($\delta\phi \ll 1$), the modes in SCET contributing to the dijet cross section are characterized by the scaling parameter $\lambda \sim \delta\phi$.
Explicitly, the momenta scale can be expressed as 
\begin{itemize}
    \item{\makebox[3.5cm]{\textbf{hard}:\hfill} $p_h^\mu \sim p_T (1,1,1),$}
    \item{\makebox[3.5cm]{$n_{a,b}$-\textbf{collinear}:\hfill} $ p_{c}^\mu \sim p_T\,(\lambda^2,1,\lambda)_{n_{i} \bar n_{i}},$}
    \item{\makebox[3.5cm]{\textbf{soft}:\hfill} $p_s^\mu\sim p_T\,(\lambda,\lambda,\lambda),$}
    \item{\makebox[3.5cm]{$n_{1,2}$-\textbf{collinear}:\hfill} $ p_{c}^\mu \sim p_T\,(\lambda^2, 1, \lambda)_{n_{i} \bar n_{i}}$.}
\end{itemize}

Since soft and collinear modes share the same virtuality, rapidity divergences arise in both sectors. These divergences cancel among the corresponding soft, beam and jet functions within the factorization theorem and can be systematically regulated and resummed through frameworks such as the Collins–Soper–Sterman (CSS) formalism~\cite{Collins:1981uk, Collins:1984kg}, the collinear anomaly approach~\cite{Becher:2010tm, Becher:2011xn}, or the rapidity renormalization group (RRG)~\cite{Chiu:2011qc, Chiu:2012ir}. In this work, we employ the RRG framework for the subsequent factorization and resummation.

As illustrated in figure~\ref{fig:coordinates}, in the limit $\delta\phi \to 0$, the azimuthal deviation is linearly related to the transverse momentum component $q_y$ via
\begin{equation}
    \delta\phi \approx \frac{q_y}{p_{T}},
\end{equation}
since $q_y = p_T \sin(\delta\phi)$. Consequently, the differential cross section for small $\delta\phi$ factorizes as 
\vspace{-0.1cm}
\begin{samepage}
\begin{align} \label{eq:fac_dphi}
\frac{\mathrm{d}^4 \sigma}{\mathrm{d} \eta_1 \mathrm{d} \eta_2 \mathrm{d} p_T \mathrm{d} q_y} = & \sum_{ijk\ell} \frac{x_a x_b}{16 \pi \hat{s}^2} \frac{2p_T}{1+\delta_{k\ell}} \bm{\mathcal H}_{ij \rightarrow k\ell, J I}(p_T, \eta_1-\eta_2, \mu) \notag\\
& \times \int_{-\infty}^{\infty} \frac{\mathrm{d} b_y}{2\pi} e^{i b_y q_y} \bm{S}_{ijk\ell, I J}(b_y, \eta_1, \eta_2, \mu, \nu) \mathcal J_k\left(b_y, \omega_1, \mu, \nu \right) \mathcal J_\ell\left(b_y, \omega_2, \mu, \nu \right) \notag\\
& \times B_{i / p}\left(x_a, b_y, \omega_a, \mu, \nu \right) B_{j / p}\left(x_b, b_y, \omega_b, \mu, \nu \right),
\end{align}
\end{samepage}
where $\omega_i$ is given by
$$\omega_i= \bar{n}_i\cdot p_i =
\begin{cases}
x_i \sqrt{s}, &i=a,b \\
2p_T \cosh{\eta_i}, &i=1,2
\end{cases}\,.
$$
In eq.~(\ref{eq:fac_dphi}), $\mu$ and $\nu$ denote the renormalization and rapidity scales, respectively. Here, the indices $i,j,k,\ell$ denote parton flavors, while the symmetry factor $1/(1+\delta_{k\ell})$ accounts for identical final-state partons. The boldface notation indicates that the hard ($\bm{\mathcal H}$) and soft ($\bm{S}$) functions are operators in color space, with subscripts $I, J$ representing color indices in the relevant representation. We note that the standard spin and color averaging factors are absorbed into the definition of the hard function \cite{Broggio:2014hoa}.

The beam functions $B_{i / p}\left(x_a, b_y, \omega_a, \mu, \nu \right)$ encode the initial-state collinear dynamics, describing the radiation emitted by partons extracted from the incoming hadrons prior to the hard interaction. By incorporating transverse recoil effects, they effectively replace the standard parton distribution functions (PDFs) in the regime of small transverse momentum. In the perturbative regime, these beam functions can be matched onto the standard collinear PDFs via an operator product expansion (OPE) as~\cite{Chiu:2012ir, Collins:1981uw, Collins:1984kg}
\begin{align}
    B_{i/p}(x, b_y, \omega, \mu, \nu) = \sum_j \int_x^1 \frac{\mathrm{d}x'}{x'} \, \mathcal{I}_{ij} \left( \frac{x}{x'}, b_y, \omega, \mu, \nu \right) f_j(x', \mu) \left[ 1 + \mathcal{O}(\Lambda_{\mathrm{QCD}}^2 \, b_T^2) \right].
\end{align}
The perturbative matching coefficients $\mathcal{I}_{ij}$ are currently known up to three-loop order~\cite{Luo:2019szz, Behring:2019quf, Ebert:2020yqt}, while the linearly polarized contributions have been determined at two-loop order~\cite{Gutierrez-Reyes:2019rug}. 

Complementarily, the jet functions capture the final-state collinear physics, governing the evolution and fragmentation of energetic partons into the observed jets. In this work, we employ the one-loop jet functions calculated in the WTA recombination scheme~\cite{Chien:2020hzh, Gutierrez-Reyes:2018qez, Gutierrez-Reyes:2019vbx, Chien:2022wiq}. We assume throughout this paper that $\delta\phi, q_T/p_{T} \ll R$, in which case the jet functions become independent of the jet radius $R$~\cite{Gutierrez-Reyes:2018qez}.

Regarding the spin structure of these sectors, we note that linearly polarized gluons can, in principle, participate in the scattering process. However, their effects are absent at the NLO accuracy of our current study, because the hard scattering only involves massless partons.

\subsubsection{Anomalous dimensions for hard, soft, beam and jet functions}
The massless hard functions are calculated to NNLO for $pp\to$ dijet process in ref.~\cite{Broggio:2014hoa}. The hard function satisfies the following RG equation,
\begin{equation}
    \frac{\df}{\df\ln{\mu}}\bm{\mathcal{H}}_{ij\to k\ell}=
    \bm{\Gamma}_{H_{ij\to k\ell}} \bm{\mathcal{H}}_{ij\to k\ell} + \bm{\mathcal{H}}_{ij\to k\ell} \bm{\Gamma}_{H_{ij\to k\ell}}^\dagger\,,
\end{equation}
where the anomalous dimension takes the form \cite{Becher:2009cu, Broggio:2014hoa} 
\begin{align}
\bm{\Gamma}_{H_{ij\rightarrow k\ell}} 
&=  \sum_{i<j } \mathbf{T}_i \cdot \mathbf{T}_j \,\gamma_{\rm cusp}(\alpha_s) \ln\frac{\mu^2}{-\hat{s}_{ij}-i0} + \gamma_H (\alpha_s)  +\mathcal{O}(\alpha_s^3) \nn \\
&= \left[\frac{C_H}{2} \gamma_{\text {cusp}}(\alpha_s) \left(\ln \frac{\hat{s}}{\mu^2}-i \pi\right)+\gamma_H (\alpha_s)\right] \bm{1}+\gamma_{\text {cusp}}(\alpha_s) \bm{M}_{ij \rightarrow k\ell} +\mathcal{O}(\alpha_s^3), 
\end{align}
with $C_H=n_q C_F+n_g C_A$ and $\gamma_H=n_q \gamma^q+n_g \gamma^g$. Here $n_q$ and $n_g$ indicate the number of quarks and gluons, respectively. 
In massless scattering, the partonic Mandelstam invariants are defined as $\hat{s}_{ij} = 2\sigma_{ij}\,p_i\cdot p_j$ ($i\neq j$), where $\sigma_{ij}=1$ if both momenta are incoming or outgoing, and $\sigma_{ij}=-1$ otherwise.
$\bm{M}_{ij \rightarrow k\ell}$ is a non-diagonal color matrix defined by the Catani–Seymour color insertion operators $\mathbf{T}_i$~\cite{Catani:1996vz} and a kinematic parameter $r = -\hat{t}/\hat{s}$, taking the form
\begin{align}
  \bm{M}_{ij \rightarrow k\ell} 
&= -(\ln r+i \pi) \left( \mathbf{T}_1 \cdot \mathbf{T}_3 + \mathbf{T}_2 \cdot \mathbf{T}_4 + \mathbf{T}_1 \cdot \mathbf{T}_4 + \mathbf{T}_2 \cdot \mathbf{T}_3 \right) \nn\\
&\quad+\ln \frac{r}{1-r} \,(\mathbf{T}_1 \cdot \mathbf{T}_4 + \mathbf{T}_2 \cdot \mathbf{T}_3). \end{align}

Analogous to the hard functions, the soft functions satisfy RG equations in color space that mix color structures under the RG evolution, 
\begin{equation}
\frac{\mathrm{d}}{\mathrm{d} \ln \mu} \bm{S}_{ijk\ell}=(\bm{\Gamma}^{S_{ijk\ell}}_\mu)^{\dagger} \bm{S}_{ijk\ell}+\bm{S}_{ijk\ell} \bm{\Gamma}^{S_{ijk\ell}}_\mu,   
\end{equation}
where the anomalous dimension reads \cite{Kidonakis:1998nf,Kidonakis:1998bk,Aybat:2006mz,Aybat:2006wq,Gao:2023ivm}
\begin{equation}
\bm{\Gamma}^{S_{ijk\ell}}_\mu = \sum_{i<j} \mathbf{T}_i \cdot \mathbf{T}_j \gamma_{\text{cusp}} (\alpha_s)\ln{\frac{ -\sigma_{ij}\nu^2 n_i \cdot n_j -i0}{2\mu^2}} - \frac{C_H}{2} \gamma^s(\alpha_s) \bm{1} +\mathcal{O}(\alpha_s^3).
\end{equation}
However, after utilizing color conservation, the rapidity divergence in the soft function is found to be proportional to the Casimir operators. This result is consistent with the cancellation of rapidity divergences between the soft function and the color-diagonal beam and jet functions. Therefore, the RRG equations of soft functions do not include any mixture of different color information, 
\begin{equation}
    \frac{\mathrm d}{\mathrm{d} \ln{\nu}} \bm{S}_{ijk\ell} = \Gamma^{S_{ijk\ell}}_\nu \bm{S}_{ijk\ell},
\end{equation}
where 
\begin{equation}
    \Gamma^{S_{ijk\ell}}_\nu = -C_H \gamma_{\text{cusp}}(\alpha_s) L_b - \left(\frac{\alpha_s}{4\pi}\right)^2 C_H \left(\frac{\beta_0}{2} \gamma_0^{\text{cusp}}L_{b}^2 -\frac{\gamma^\nu_1}{2} \right) + \mathcal{O}(\alpha_s^3),
\end{equation}
and the first term contains both an $\alpha_s$ and $\alpha_s^2$ contribution,
with $L_b = \ln{(\mu^2 b^2/b_0^2)}$ and $b_0 = 2e^{-\gamma_E}$. 

By investigating the anomalous dimensions of beam functions and jet functions, we find that for $\mathcal{F}_i\in\{B_i, \, \mathcal{J}_i\}$, their RG equations are as follows,
\begin{equation}
    \frac{\mathrm d}{\mathrm d \ln\mu} \mathcal{F}_i(b,\mu,\nu) = \Gamma^{\mathcal{F}_i}_\mu(b,\nu)~\mathcal F_i(b, \mu, \nu),
\end{equation}
where 
\begin{equation}
    \Gamma^{\mathcal{F}_i}_\mu(b,\nu) = 2C_i\gamma_{\text{cusp}}(\alpha_s)\ln{\frac{\nu}{\omega_i}} + \gamma^{\mathcal{F}_i}.
\end{equation}

RRG equations for $\mathcal{F}_i$ take the form
\begin{equation}
    \frac{\mathrm d}{\mathrm d \ln\nu} \mathcal{F}_i(b,\mu,\nu) = \Gamma^{i}_\nu(b,\mu) ~ \mathcal F_i(b, \mu, \nu),
\end{equation}
where
\begin{equation} \label{eq:RRG_AD}
    \Gamma^{i}_\nu(b, \mu) = C_i \gamma_{\text{cusp}} L_b + \left(\frac{\alpha_s}{4\pi}\right)^2 C_i \left(\frac{\beta_0}{2} \gamma_0^{\text{cusp}}L_{b}^2 -\frac{\gamma^\nu_1}{2} \right) + \mathcal{O}(\alpha_s^3).
\end{equation}
Explicit expressions for the perturbative expansions of the non-cusp anomalous dimensions $\gamma^q$, $\gamma^g$, $\gamma^s$, $\gamma^{\mathcal{F}_i}$ and the non-cusp rapidity anomalous dimension $\gamma^\nu$ are provided in appendix~\ref{app:anomalous}. 

As a non-trivial check of our theoretical framework, we have verified the RRG consistency 
\begin{align}
    \Gamma_\nu^{S_{ijkl}} + \sum_{i=a,b,1,2} \Gamma_\nu^i = 0\,,
\end{align}
and the RG consistency
\begin{align}
    \frac{\df}{\df\ln\mu} \left[\tr(\bm{\mathcal{H}} \bm{S}) B_a B_b \mathcal{J}_1 \mathcal{J}_{2} \right] = 0\,.
\end{align}

\subsection{Transverse momentum imbalance} \label{sec:qT_fact}
In SCET, the kinematic modes relevant for the $q_T$ distribution are given by
\begin{itemize}
    \item{\makebox[3.5cm]{\textbf{hard}:\hfill} $p_h^\mu \sim p_T (1,1,1),$}
    \item{\makebox[3.5cm]{$n_{a,b}$-\textbf{collinear}:\hfill} $ p_{c_i}^\mu\sim (q_T^2/p_T, p_T, q_T)_{n_i \bar n_i} $,}
    \item{\makebox[3.5cm]{$n_{1,2}$-\textbf{collinear}:\hfill} $ p_{c_i}^\mu\sim (q_y^2/p_T, p_T, q_y)_{n_i \bar n_i}$,}
\end{itemize}
where the soft modes are not shown here, as their structure is more intricate and requires further refactorization in the small-$R$ limit.

The factorization formula for the $q_T$ distribution is written as 
\begin{align} \label{eq:fac_qT_mom}
\frac{\mathrm{d}^4 \sigma}{\mathrm{d} \eta_1 \mathrm{~d} \eta_2 \mathrm{~d} p_T \mathrm{~d} q_T} =& \sum_{ijk\ell} \frac{x_a x_b}{16 \pi \hat{s}^2} \frac{2p_T}{1+\delta_{k\ell}} 
~q_T\int_0^{2\pi} \!\! \df \phi_q \int \df\vec{k}_{s,T}\, \df\vec{k}_{a,T}\, \df\vec{k}_{b,T}\, \df k_{1,y}\, \df k_{2,y} \\
&\times 
\tr\left[\bm{\mathcal H}_{ij \rightarrow k\ell}(p_T, \eta_1-\eta_2, \mu)\bm{S}_{ijk\ell}(\vec{k}_{s,T}, \eta_1, \eta_2, R,\mu,\nu) \right]   \nn\\
&\times\mathcal J_k\left(k_{1,y}, \omega_1, \mu, \nu \right) \mathcal J_\ell\left(k_{2,y}, \omega_2, \mu, \nu \right) \nn\\
&\times B_{i / p}(x_a, \vec{k}_{a,T}, \omega_a, \mu, \nu )\, B_{j / p}(x_b, \vec{k}_{b,T}, \omega_b, \mu, \nu) \nn\\
&\times \delta(q_y + k_{s,y} + k_{a,y} + k_{b,y} + k_{1,y} + k_{2,y})\, \delta(q_x + k_{s,x} + k_{a,x} + k_{b,x})\,. \nn
\end{align}
The soft function in eq.~\eqref{eq:fac_qT_mom} differs from the $q_y$ soft function. In particular, it distinguishes radiation inside and outside jets and thus depends on the jet radius.

We can apply a Fourier transform to the factorization ingredients in eq.~\eqref{eq:fac_qT_mom}, converting the convolution of different $\vec{k}_{i,\,T}$ into a product in $\vec{b}_T$ space,
\begin{align}
    \bm{S}_{ijk\ell}(\vec{b}_{T}, \eta_1, \eta_2, R,\mu,\nu) &= \int_{-\infty}^\infty \df\vec{k}_{s,T} ~e^{i\vec{k}_{s,T} \cdot \vec{b}_T}~
    \bm{S}_{ijk\ell}(\vec{k}_{s,T},\eta_1, \eta_2, R,\mu,\nu)\,, \\
    B_{i / p}(x_a, \vec{b}_{T}, \omega_a, \mu, \nu ) &= \int_{-\infty}^\infty \df\vec{k}_{a,T} ~e^{i\vec{k}_{a,T}\cdot \vec{b}_T}
    B_{i / p}(x_a, \vec{k}_{a,T}, \omega_a, \mu, \nu )\,, \\
    J_k\left(b_y, \omega_1, \mu, \nu \right) 
    &= \int_{-\infty}^{\infty} \df k_{1,y} ~e^{ik_{1,y} b_y} J_k(k_{1,y}, \omega_1, \mu, \nu)\,.
\end{align}
The hard, jet, and beam functions entering here are defined in the same way as in the $\delta\phi$ distribution. The only modification arises in the TMD beam functions: here they depend on the full transverse momentum of the initial collinear modes rather than solely on the $y$-component. Fortunately, since TMD beam functions are isotropic in the transverse plane, the replacement $b_y \to b_T$ in $B_{i/p}(x_a,b_y, \omega_a, \mu,\nu)$ suffices.
The most significant difference between the factorization formulas for the $\delta\phi$ and $q_T$ distributions lies in the soft function. 
The soft function $\bm{S}_{ijk\ell}(\vec{b}_{T}, \eta_1, \eta_2, R,\mu,\nu)$ describes the recoil from soft radiation emitted by the beams and jets on the two WTA axes. We will elaborate on its detailed refactorization structure in the small-$R$ limit in section~\ref{sec:soft_refact}.

The factorization formula for the $q_T$ spectrum, written in impact parameter space, is given by
\begin{align} \label{eq:fac_qT}
\frac{\mathrm{d}^4 \sigma}{\df\eta_1\, \df\eta_2\, \df p_T\, \df q_T}= & \sum_{ijk\ell} \frac{x_a x_b}{16 \pi \hat{s}^2} \frac{2p_T}{1+\delta_{k\ell}} \bm{\mathcal H}_{ij \rightarrow k\ell, J I}(p_T, \eta_1-\eta_2, \mu) ~q_T
\int_0^{2\pi} \!\! \df \phi_q
\int\! \frac{\mathrm{d}^2 \vec{b}_T}{(2\pi)^2} e^{i\vec{q}_T\cdot \vec{b}_T}~
\notag\\
& \times  \bm{S}_{ijk\ell, I J}(\vec{b}_T, \eta_1, \eta_2, R, \mu, \nu) \mathcal J_k\left(b_y, \omega_1, \mu, \nu \right) \mathcal J_\ell\left(b_y, \omega_2, \mu, \nu \right) \notag\\
& \times B_{i / p}(x_a, \omega_a, b_T, \mu, \nu )\, B_{j / p}(x_b, \omega_b, b_T, \mu, \nu).
\end{align}

\subsection{Refactorization of \texorpdfstring{$q_T$}{qT}-soft function} \label{sec:soft_refact}
\begin{figure}[tp]
    \centering
    \includegraphics[width=0.8\linewidth]{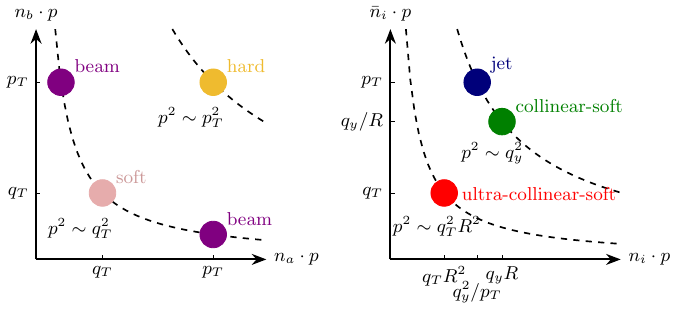}
    \caption{Schematic representation of the relevant modes in the $q_T$ factorization for $pp\to$ dijets in the small-$R$ limit.
    \textbf{Left:} Standard modes (hard, beam, and soft), where the soft mode represents the global soft function. 
    \textbf{Right:} Modes required in the small-$R$ limit, showing the separation of the standard collinear sector into jet, collinear-soft, and ultra-collinear-soft modes. Here $n_i = n_1$ or $n_2$. 
    }
    \label{fig:fact_modes}
\end{figure}

In the small-$R$ limit\footnote{Note that this does not contradict the assumptions made for the jet function, as the two cases are compatible in the intermediate regime $q_T/p_{T}\ll R \ll 1$.}, the TMD soft function can be refactorized into the product of a global TMD soft function, which is independent of the jet definition, and the functions, $S_i$, describing soft radiation near the jet boundaries. The relevant modes for this factorization are
\begin{itemize}
    \item{\makebox[4.5cm]{\textbf{soft}:\hfill} $p_s^\mu \sim q_T \,(1,1,1)$},
    \item{\makebox[4.5cm]{\textcolor{green!50!black}{$n_{i}$-\textbf{collinear-soft}}:\hfill} $ p_{cs}^\mu \sim |q_y|/R\,(R^2,1,R)_{n_i \bar n_i} $},
    \item{\makebox[4.5cm]{{\color{red}$n_{i}$-\textbf{ultra-collinear-soft}}:\hfill} $ p_{ucs}^\mu \sim q_T \,(R^2,1,R)_{n_i \bar n_i}$} ,
\end{itemize}
with $i=1$ or $2$. The soft modes share the same virtuality, $q_T^2$, as the beam functions, while the collinear-soft modes and jet functions have the same virtuality, $q_y^2$, as illustrated in figure~\ref{fig:fact_modes}. We employ the RRG framework to address the pairwise cancellation of rapidity divergences between modes with the same virtuality.

Based on this mode analysis, we factorize the small-$R$ $q_T$-soft function into 
\begin{align}
&\bm{\mathcal{S}}_{ij k\ell}(\vec{b}_T,R, \phi_1, \phi_2, \eta_1, \eta_2, \mu, \nu) = \nn\\
&\quad \bm{\mathcal{S}}_{ij k\ell}^{\rm global}(\vec{b}_T, \phi_1, \phi_2, \eta_1, \eta_2, \mu, \nu) S_{k}(\vec{b}_T, \phi_1, \eta_1, R,\mu,\nu) S_{\ell}(\vec{b}_T, \phi_2, \eta_2, R,\mu,\nu),
\end{align}
where the directions $n_i$ are here specified by their rapidity $\eta_i$ and azimuthal angle $\phi_i$.
We choose our coordinates such that the azimuthal angle $\phi_1 = 0$ for the final-state parton~1 and $\phi_2 = \pi$ for the final-state parton~2. For brevity, we will suppress this azimuthal dependence in the arguments of the soft functions from here on.
The global soft function $\bm{\mathcal{S}}_{ijk\ell}^{\rm global}$ encodes soft radiation from both the incoming beams and the outgoing jets and is independent of the jet definition details. It depends on the jet directions, as the corresponding Wilson lines source soft radiation, but only the total transverse momentum of the global soft radiation is probed through the $q_T$ measurement.

The functions $S_{i}$ encode the contributions from the $n_i$-collinear-soft and $n_i$-ultra-collinear-soft modes. Consequently, $S_i$ is not a single scale function and requires further refactorization. This function describes soft emissions near the jet boundary, which are subject to NGLs beginning at NNLO. To simplify the discussion that follows, we use a single leg index $i \in \{1,2\}$ to denote both the flavor index ($k$ or $\ell$) and the corresponding kinematic index ($1$ or $2$).

Inside the jet cone, only the $y$-component of the soft radiation is measured, whereas outside the jet cone, the full transverse momentum is observed. In the small-$R$ limit, the out-of-cone region technically occupies nearly the entire phase space. However, as demonstrated by the method-of-regions analysis in ref.~\cite{Fu:2024fgj}, the wide-angle soft radiation away from the jet boundary is scaleless; the real and virtual corrections cancel exactly, resulting in a vanishing contribution to the $q_T$ soft function. 
Consequently, the non-zero out-cone contribution arises exclusively from the ultra-collinear-soft radiation localized near the jet boundary. For the momentum $k$ of such a mode, we have 
\begin{align} \label{eq:k_order}
    k^- \gg k^+,\, k_\perp,
\end{align}
where we have adopted the light-cone coordinates along the jet axis defined in eq.~\eqref{eq:lightcone_comp}.
Under this hierarchy, the Fourier exponent associated with the out-of-cone measurement admits the power expansion 
\begin{align}
    \underbrace{\vec{b}_T \cdot \vec{k}_T}_{\text{beam coordinates}} = -b \cdot k \approx \underbrace{- \frac{1}{2} b^+ k^-}_{\text{jet coordinates}},
\end{align}
with $b^\mu = (0, b_x, b_y, 0)$ and $b^+ = - b_x n_{i,x}$.

\begin{figure}[t]
    \centering
    \includegraphics[width=0.65\linewidth]{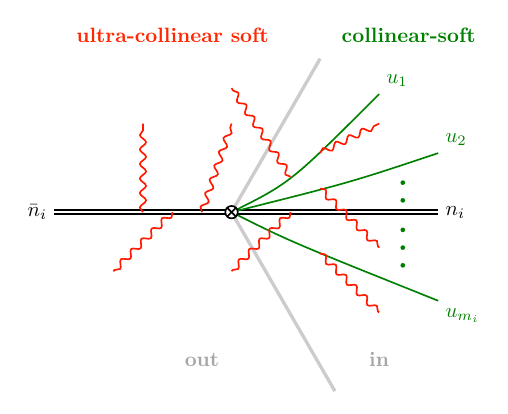}
    \caption{Pictorial representation of the factorization formula for the function $S_i$. The green lines correspond to the collinear-soft radiation, and the red lines represent ultra-collinear-soft emissions. Although the ultra-collinear-soft radiation contributes to the measurement only when it falls outside the jet cone, the radiation itself is not geometrically constrained and can also occur inside the jet.}
    \label{fig:refactorization}
\end{figure}

Applying the factorization framework for non-global observables \cite{Becher:2015hka, Becher:2016mmh, Becher:2016omr} (see also \cite{Larkoski:2015zka, Caron-Huot:2015bja}), $S_i$ can be expressed as
\begin{align} \label{eq:soft_refact1}
    &S_i (\vec{b}_T, \eta_i, R,\mu,\nu) =  \notag \\
    &\hspace{0.5cm}\sum_{m_i=0}^\infty \prod_{j=1}^{m_i} \int  \frac{\df\Omega({\vec{u}_{j}})}{4 \pi} \frac{1}{d_i}{\rm Tr}\left[ \bm{\mathcal{S}}_{m_i}^{\rm cs}(\{n_i,\bar n_i, \underline{u}\}, b_y, R,\mu,\nu)  \bm{\mathcal{S}}_{m_i}^{\rm ucs}(\{n_i,\bar n_i, \underline{u}\}, b_x, R,\mu) \right].
\end{align}
Here, both the collinear-soft and ultra-collinear-soft functions are operators in color space, as indicated by the boldface notation, and therefore necessitate a color trace ${\rm Tr}[\cdots]$. The normalization factor $d_i$ is $N_c$ for quark jets and $N_c^2-1$ for gluon jets. 
They are also (diagonal) operators in the space of collinear parton multiplicities. The collinear-soft function, $\bm{\mathcal{S}}_{m_i}^{\rm cs}$, represents the squared amplitude for $m_i$-parton emissions from the Wilson lines along $n_i$ and $\bar n_i$ into the jet region. It is integrated over emission energies while holding the directions $\{\underline{u}\}\equiv\{u_1, \dots, u_{m_i}\}$ fixed, where $u_j^\mu=(1,\vec{u}_j)$ are light-like vectors.
$\Omega({\vec{u}_{j}})$ denotes the solid angle of the three-vector $\vec{u}_{j}$. The ultra-collinear-soft function, $\bm{\mathcal{S}}_{m_i}^{\rm ucs}$, consists of $m_i+2$ Wilson lines along the directions $\{\underline{u}\}$ of the $m_i$ collinear-soft partons, the jet direction $n_i$, and the auxiliary light-like vector $\bar n_i$. The pictorial representation of the factorization is given in figure~\ref{fig:refactorization}. 

For notational simplicity, we use $\braket{\cdots}$ to denote the normalized color trace $\frac{1}{d_i}\mathrm{Tr}[\cdots]$. Furthermore, we introduce the symbol $\otimes$ to represent integration over the $m_i$ directions of the collinear-soft emissions $\{\underline{u}\}$, which allows for a compact rewriting of $S_i$:
\begin{align}
    S_i (\vec{b}_T, \eta_i, R,\mu,\nu) =  
    &\sum_{m_i=0}^\infty \left\langle \bm{\mathcal{S}}_{m_i}^{\rm ucs}(\{n_i,\bar n_i, \underline{u}\}, b_x, R,\mu) \otimes \bm{\mathcal{S}}_{m_i}^{\rm cs}(\{n_i,\bar n_i, \underline{u}\}, b_y, R,\mu,\nu) \right\rangle.
\end{align}

\subsubsection{Collinear-soft and ultra-collinear soft functions} 
We follow the notation of ref.~\cite{Becher:2016omr} to study the multi–Wilson-line factorization framework for the collinear-soft and ultra-collinear-soft functions in the small-$R$ limit. 

The amplitudes for the emission of $m_i$ collinear-soft partons with momenta $\{\underline{p}\}$ from the back-to-back Wilson lines along $n_i$ and $\bar{n}_i$ are expressed as  
\begin{align}
    \left| \mathcal{M}_{m_i}^{cs} (n_i, \bar{n}_i, \{\underline{p}\})\right\rangle =  \langle\{\underline{p}\}|S(n_i) S^\dagger (\bar{n}_i) |0\rangle,
\end{align}
where $|\{\underline{p}\}\rangle = \prod_{k=1}^{m_i} |p_k\rangle$ denotes the Hilbert space state of ${m_i}$ partons, with momenta $p_k = E_k u_k$ and energies $E_k$, and $S(n_i)$ represents the collinear-soft Wilson line associated with jet~$i$.
The amplitude $\left| \mathcal{M}_{m_i}^{cs} (n_i, \bar{n}_i, \{\underline{p}\}) \right\rangle$ is a vector in color space~\cite{Catani:1996jh, Catani:1996vz}; hence, we employ the bra--ket notation to make this structure explicit.

In the small-$R$ limit, since the virtuality of the collinear-soft mode is much larger than that of the ultra-collinear-soft mode, one can separate the two submodes and dress the collinear-soft field with the ultra-collinear-soft Wilson line $U_i(n_i)$, 
\begin{align}
    U_i(n_i) = \mathbf{P} \exp\left( i g_s \int_0^\infty \df t\, n_i \cdot A_{ucs}^a(t n_i)\, \mathbf{T}_i^a \right),
\end{align}
where $\mathbf{T}_i$  encodes the color-representation information of parton~$i$. The symbol $\mathbf{P}$ indicates path ordering of the color matrices, ensuring that the matrix-valued field $A_{ucs}^a(t n_i)$ with the larger value of $t$ appears to the left of those with smaller~$t$.
At the amplitude level, the ultra-collinear-soft radiation emitted from the two original Wilson lines together with the additional collinear-soft partons is described by the Wilson-line operator 
\begin{align}
    U_a(n_i)U_b(\bar{n}_i) U_1(u_1)\cdots U_{m_i}(u_{m_i}) \left| \mathcal{M}_{m_i}^{cs} (n_i, \bar{n}_i, \{\underline{p}\})\right\rangle.
\end{align}

Now we can define the collinear-soft function in $\vec{b}_T$ space as 
\begin{align} \label{eq:cs_def}
    &\bm{\mathcal{S}}_{m_i}^{\rm cs}(\{n_i,\bar n_i, \underline{u}\}, b_y, R,\mu,\nu)  =   \\
    &\hspace{0.5cm}
    \prod_{j=1}^{m_i} \int \frac{\df E_j E_j^{d-3}}{(2\pi)^{d-2}}  \left| \mathcal{M}_{m_i}^{cs} (n_i, \bar{n}_i, \{\underline{p}\})\right\rangle  \left\langle \mathcal{M}_{m_i}^{cs} (n_i, \bar{n}_i, \{\underline{p}\})\right| 
    \exp \left(ib_y\sum_{k=1}^{m_i} p_{k,y}\right) \Theta_{\rm in}^i(\{\underline{p}\}),\notag
\end{align}
Here, the phase-space constraint $\Theta_{\rm in}^i(\{\underline{p}\})$ restricts all $m_i$ collinear-soft partons emitted from the original Wilson lines to lie inside the jet cone labeled with $i$. As the refactorization is carried out in the small-$R$ limit, the in-cone constraint simplifies to
\begin{align}
    \Theta_{\rm in}^i(\{\underline{p}\}) = \prod_{j=1}^{m_i} 
    \Theta \biggl(R_i^2 - \frac{n_i\cdot p_j}{\bar{n}_i\cdot p_j}\biggr), \quad R_i = \frac{R}{2\cosh{\eta_i}}.
\end{align}

The ultra-collinear-soft function in momentum space can be written as
\begin{align} \label{eq:ucs_def}
    \bm{\mathcal{S}}_{m_i}^{\rm ucs}(\{n_i,\bar n_i, \underline{u}\}, b_x, R,\mu) &= \SumInt_
    {X_{ucs}} \langle0|    U^\dagger_{m_i}(u_{m_i}) \cdots U^\dagger_1(u_1) U^\dagger_b(\bar{n}_i) U^\dagger_a(n_i)| X_{ucs} \rangle  \\
    &\!\! \langle X_{ucs} | U_a(n_i)U_b(\bar{n}_i) U_1(u_1)\cdots U_{m_i}(u_{m_i}) | 0\rangle \exp \biggl( -\frac{i}{2}b^+\sum_{j\notin \mathrm{jet-}i} k_{j}^{-} \biggr).\notag
\end{align}
where $| X_{ucs} \rangle = \prod_j |k_j \rangle$, and the measurement function in the Fourier exponent captures only the unresolved ultra-collinear-soft partons outside the jet cone.

Details regarding the renormalization and anomalous dimensions of the collinear-soft and ultra-collinear-soft functions in multiplicity space are provided in appendix~\ref{sec:renorm_cs_ucs}.

\section{Resummation} \label{sec:resum}

\subsection{Azimuthal decorrelation}
To resum all the large logarithms, we evolve the hard function from its natural scale $\mu_h$ to the factorization scale $\mu_f$, which we choose equal to the natural scale for the soft and collinear sectors. We also evolve the rapidity scales of jet functions and beam functions from their respective characteristic rapidity scales $\nu_i = \omega_i$ to the soft function's rapidity scale $\nu_s = b_0/|b_y|$ with $b_0 = 2e^{-\gamma_E}$.
This allows us to derive the resummation formula
\begin{align} \label{eq:resum_dphi}
\frac{\mathrm{d}^4 \sigma}{\mathrm{d} \eta_1 \mathrm{~d} \eta_2 \mathrm{~d} p_T \mathrm{~d} q_y} = & \sum_{ijk\ell} \frac{x_a x_b}{16 \pi \hat{s}^2} \frac{2p_T}{1+\delta_{k\ell}} \int_{0}^{\infty} \frac{\mathrm{d} b_y}{\pi} \cos \left(b_y q_y \right) 
\prod_{i=a,b,1,2}\left(\frac{\nu_s}
{\nu_i}\right)^{\Gamma^{i}_\nu(b_y,\, \mu_f)} \notag \\
& \times \sum_{KK'} \exp\left\{ \int_{\mu_h}^{\mu_f} \frac{\mathrm d \mu}{\mu} \left[\gamma_{\text{cusp}}(\alpha_s)  \left( C_H \ln{\frac{\hat{s}}{\mu^2}} + \lambda_K + \lambda_{K'}^* \right) + 2 \gamma_H(\alpha_s) \right]\right\}  \notag \\
& \times \bm{\mathcal H}_{ij \rightarrow k\ell, KK'}(p_T, \eta_1-\eta_2, \mu_h)  \bm{S}_{ijk\ell, K'K}(b_y, \eta_1, \eta_2, \mu_f, \nu_s)  \notag \\
& \times B_{i / p}\left(x_a, \omega_a, b_y, \mu_f, \nu_a \right) B_{j / p}\left(x_b, \omega_b, b_y, \mu_f, \nu_b \right) \nn\\
&\times\mathcal J_k\left(b_y, \omega_1, \mu_f, \nu_1 \right) \mathcal J_\ell\left(b_y, \omega_2, \mu_f, \nu_2 \right) \notag\\
& \times \exp\left[-S^i_{\text{NP}}(b_y,Q_0,\omega_a) -S^j_{\text{NP}}(b_y,Q_0,\omega_b) \right],
\end{align}
where $\lambda_K$ are the eigenvalues of $\bm{M}_{ij\to k\ell}$ in the hard function's anomalous dimensions.
Here, $\mu_h = v_h \, 2 p_T$ and $\mu_f = \min\{v_f \, \mu_{b_{y,*}},\,\mu_h\}$, where $v_h$ and $v_f$ are scale-variation parameters used to estimate theoretical uncertainties. 
Our central curves in section~\ref{sec:numerical} are obtained with $v_h = v_f = 1$, and  uncertainty bands correspond to varying $v_h, v_f \in [0.5, 2]$.

To avoid the Landau pole at large~$b_y$, we adopt the $b_*$ prescription,
\begin{align}
    b_{y,*} = \frac{|b_y|}{\sqrt{1+b_y^2/b_{\rm max}^2}}, \quad \mu_{b_{y,*}} = \frac{b_0}{b_{y,*}},
\end{align}
with $b_{\max} = 1.5~\text{GeV}^{-1}$. To test the robustness of our calculation we varied $b_{\max}$ around this value by a factor of two, which yields negligible variations in the resummed predictions.
To account for non-perturbative effects that dominate at large impact parameters $b$, we introduce a non-perturbative Sudakov form factor, $S_{\rm NP}$ \cite{Landry:2002ix, Konychev:2005iy, Collins:2014jpa, Aidala:2014hva, Sun:2014dqm}.  
Specifically, its explicit form is given by 
\begin{equation} \label{eq:SNP}
    S_{\rm NP}^{i}(b,Q_0,Q) = g_1 b^2 + \frac{g_2\,C_{i}}{2\,C_F}\ln\frac{Q}{Q_0} \ln{\frac{b}{b_*}},
\end{equation}
where the first term parametrizes the Gaussian shape and the second term captures the nonperturbative evolution. Here, $C_{i}$ is the quadratic Casimir. For the numerical evaluation, we employ the fitting parameters at $Q_0^2 = 2.4~\mathrm{GeV}^2$ obtained from  ref.~\cite{Sun:2014dqm}: $g_1=0.106~\mathrm{GeV}^{2}$, $g_2=0.84$.

\subsection{Transverse momentum imbalance} \label{sec:qT_resum}
In the subsequent resummation, performed in the conjugate $\vec{b}_T$-space, the soft (global soft function) and initial collinear (beam function) modes are assigned a common scale $\mu_s \sim \mu_b$, while the collinear-soft and final collinear (jet function) modes share the scale $\mu_{cs} \sim \mu_j$. The ultra-collinear-soft mode is characterized by its own scale $\mu_{ucs}$. Their typical values are
 \begin{align}
     \mu_b\sim \mu_s\sim b_0/b_T,\quad \mu_{cs} \sim \mu_j \sim b_0/|b_y|,\quad \mu_{ucs} \sim R\, b_0/ (2|b_x|),
 \end{align}
 together with the hard scale $\mu_h \sim 2p_T$. 
We again choose the factorization scale as the common scale of the beam and the global-soft functions, $\mu_f \sim b_0/b_T$. Since the two groups of modes share common virtualities within each group, we employ the RRG framework to handle the associated rapidity divergences~\cite{Chiu:2011qc, Chiu:2012ir}. The characteristic rapidity scales for different modes are given by
\begin{align}
    \nu_{i} \sim \omega_i \,(i=a,b,1,2), \quad \nu_{s} \sim b_0/b_T, \quad \nu_{cs,i} \sim b_0/(|b_y|R_i) \,(i=1,2).
\end{align}

It is also worth noting that though the virtualities of the soft and collinear-soft modes are parametrically of the same order, their numerical values are not identical. Therefore, the renormalization group (RG) evolution from $\mu_{cs}$ to $\mu_f$ should be included to resum the large logarithms $\ln(|b_y|/b_T)$ that arise in the limit $\sin\phi_b \to 0$. 

After analyzing the modes and scales, we proceed with the RRG and RG evolution.
The rapidity scales of the beam functions are evolved from $\nu_{i} = x_i \sqrt{s}$ ($i = a, b$) to the rapidity scale of the global soft function, $\nu_{s} = b_0/b_T$.
Similarly, the rapidity scales of the jet functions should first be evolved from $\nu_{i} = 2p_T \cosh{\eta_i}$ ($i = 1, 2$) to the rapidity scale of the collinear-soft function, $\nu_{cs,i} = b_0/(|b_y|R_i)$. The rapidity scale evolution takes the form of
\begin{align}
&  \prod_{i=ab}\! \left(\frac{\nu_{s}} {\nu_{i}}\right)^{\!\Gamma^{i}_\nu(b_{T},\, \mu_f)}
    \prod_{i=12}\! \left(\frac{\nu_{cs,i}} {\nu_{i}}\right)^{\!\Gamma^{i}_\nu(b_y,\, \mu_{cs})}
= \prod_{i=ab}\! \left(\frac{b_0} {b_T x_i \sqrt{s}} \right)^{\!\Gamma^{i}_\nu(b_{T},\, \mu_f)}
    \prod_{i=12}\! \left(\frac{b_0} {|b_y|p_T R}\right)^{\!\Gamma^{i}_\nu(b_y,\, \mu_{cs})}.
\end{align}

Next, we consider the RG evolution to the factorization scale. The RG evolution of the hard function is analogous to that for the $q_y$ distribution, while the jet, collinear-soft, and ultra-collinear-soft functions require additional clarification. The evolution matrix for these functions can be schematically written as 
\begin{align}
    U &= \exp{\left[\int_{\mu_j}^{\mu_f} \frac{\df\mu}{\mu} \Gamma^{J_i}\right]}\,
    \mathbf{P}\exp{\left[\int_{\mu_{cs}}^{\mu_f} \frac{\df\mu}{\mu} \bm{\Gamma}^{\rm cs} - \int_{\mu_{ucs}}^{\mu_f} \frac{\df\mu}{\mu} \bm{\Gamma}^{\rm ucs}\right]} \nn\\
    &= \exp{\left[\int_{\mu_{cs}}^{\mu_f} \frac{\df\mu}{\mu} \left(\Gamma^{J_i} + \Gamma^{\rm cs} + \Gamma^{\rm ucs} \right) \right]}\,
    \mathbf{P}\exp{\left[- \int_{\mu_{ucs}}^{\mu_{cs}} \frac{\df\mu}{\mu} \bm{\Gamma}^{\rm ucs}\right]},
\end{align}
where the jet scale $\mu_{j}$ is the same as the collinear-soft scale $\mu_{cs}$. The separation of the anomalous dimensions for $\bm{\Gamma}^{\rm{cs},\,\rm{ucs}}$ in multiplicity space is defined in eqs.~\eqref{eq:separate_ad_cs} and \eqref{eq:separate_ad_ucs}, while the explicit expressions for their global parts, $\Gamma^{\rm{cs},\,\rm{ucs}}$, are given in eqs.~\eqref{eq:ad_cs} and \eqref{eq:ad_ucs}.

After combining the evolution matrices for the collinear-soft and ultra-collinear-soft functions from $\mu_{cs}$ to the factorization scale $\mu_f$, the multiplicity mixing arises only in the second exponential term of the following evolution
\begin{align}
    \exp{\left[ - \int_{\mu_{ucs}}^{\mu_{cs}} \frac{\df\mu}{\mu} \bm{\Gamma}^{\rm ucs}\right]}
    = \exp{\left[\int_{\mu_{ucs}}^{\mu_{cs}} \frac{\df\mu}{\mu}\Gamma^{\rm ucs}\right]} \, \mathbf{P}\exp{\left[- \int_{\mu_{ucs}}^{\mu_{cs}} \frac{\df\mu}{\mu} \bm{\hat\Gamma}\right]}.
\end{align}

We can define the multiplicity-mixing evolution matrix in a more rigorous manner. 
To resum the leading NGLs $\ln R$ (denoted as LL$_R$), one can retain only the tree-level contribution of the ultra-collinear-soft function, $\bm{\mathcal{S}}_{m}^{\rm{ucs}} = \bm{1} + \mathcal{O}(\alpha_s)$, together with the leading-order collinear-soft function, $\bm{\mathcal{S}}_{0}^{\rm cs} = \bm{1}$, to evaluate 
\begin{align}
    U_{\rm NG}^i (\mu_{ucs}, \mu_{cs}) \stackrel{\text{LL}_R}{=} \sum_{m=0}^\infty \left\langle \bm{1} \,\hat{\otimes}\, \bm{U}_{m0}(\{n_i,\bar n_i, \underline{u}\}, \mu_{ucs}, \mu_{cs}) \right\rangle,
\end{align}
where $\bm{U}(\{n_i, \bar{n}_i, \underline{u}\}, \mu_{ucs}, \mu_{cs}) \equiv \mathbf{P}\exp{\left[- \int_{\mu_{ucs}}^{\mu_{cs}} \df\ln\mu\, \bm{\hat\Gamma}(\{\underline{n}\}, \mu) \right]}$.

As shown in the appendix \ref{app:hemisphere}, the ultra-collinear-soft and collinear-soft functions defined here can be interpreted as hemisphere soft functions after performing a Lorentz boost along the jet axis, analogous to the constructions in \cite{Becher:2016mmh, Dasgupta:2012hg}.
Following the procedure outlined in \cite{Chien:2019gyf}, the non-global evolution matrix from the ultra-collinear-soft scale to the collinear-soft scale can be approximated using the Dasgupta–Salam parametrization~\cite{Dasgupta:2001sh},
\begin{align} \label{eq:U_NGL}
    U_{\rm NG}^i (\mu_{ucs}, \mu_{cs}) \approx \exp\left( -C_A C_i \frac{\pi^2}{3} u^2 \frac{1 + (a u)^2}{1 + (b u)^c} \right), \quad
    u =  \frac{1}{\beta_0} \ln \frac{\alpha_s(\mu_{ucs})}{\alpha_s(\mu_{cs})},
\end{align}
with the constants given by $a=0.85C_A$, $b=0.86C_A$ and $c=1.33$. One can resum the leading NGL contributions using eq.~\eqref{eq:U_NGL} in the subsequent NLL and NNLL resummations.

Therefore, the combined RG evolution for the jet, collinear-soft and ultra-collinear-soft functions is given by
\begin{align}
    \prod_{i=1,2}\exp{\left[\int_{\mu_{cs} }^{\mu_f} \frac{\df\mu}{\mu} \left(\Gamma^{J_i} + \Gamma^{\rm cs}_i \right) 
    +  \int_{\mu_{ucs} }^{\mu_f} \frac{\df\mu}{\mu}  \Gamma^{\rm ucs}_i \right]} 
    U_{\rm NG}^i (\mu_{ucs}, \mu_{cs}).
\end{align}
After evolving the rapidity scales of the jet functions, $\nu_i$ $(i=1,2)$, to the rapidity scale of the corresponding collinear-soft functions $\nu_{cs,i}$ $(i=1,2)$, we can combine the jet and collinear-soft contributions and perform their RG evolution together, yielding
\begin{align}
    U^{\text{jet}+cs}(\mu_{cs}, \mu_f) 
    &= \prod_{i=1,2}\exp{\left[\int_{\mu_{cs} }^{\mu_f} \frac{\df\mu}{\mu} \left(\Gamma^{J_i} + \Gamma^{\rm cs}_i \right) \right]} \\
    &= \prod_{i=1,2} \exp \left\{ \int_{\mu_{cs}}^{\mu_{f}} \frac{\df\mu}{\mu} \sum_{i=1}^{2} \left[ C_i \gamma_{\text{cusp}} \ln \frac{\mu^2}{p_T^2R^2} + \gamma_{nc,i} \right] \right\} \notag\\
    &= \prod_{i=1}^{2} \left[ \exp \left( -2 C_i S(\mu_{cs}, \mu_{f}) - A_{\gamma_{nc,i}} (\mu_{cs}, \mu_{f}) \right) \left( \frac{p_T^2R^2}{\mu_{cs}^2} \right)^{C_i A \gamma_{\text{cusp}}(\mu_{cs}, \mu_{f})} \right], \nn
\end{align}
where the cusp $\gamma_{\text{cusp}}$ and the non-cusp anomalous dimension $\gamma_{nc,i} = \gamma^{J_i} + \mathcal{O}(\alpha_s^3)$ are given in appendix~\ref{app:anomalous}. The perturbative expression of $S(\nu,\mu)$ and $A_{\gamma}(\nu,\mu)$ are shown in the appendix of \cite{Becher:2006mr}.
As for the global evolution of the ultra-collinear-soft function, it takes the form
\begin{align} \label{eq:Uucs_global}
    U^{ucs}(\mu_{ucs}, \mu_{f}) &= \prod_{i=1,2}\exp{\left[ \int_{\mu_{ucs} }^{\mu_f} \frac{\df\mu}{\mu}  \Gamma^{\rm ucs}_i \right]} \nn \\
    &= \prod_{i=1,2} \exp{\left[ 2 C_i S(\mu_{ucs}, \mu_{f}) -i\pi \,C_i A_{\gamma_{\rm csup}} (\mu_{ucs}, \mu_{f})\, \mathrm{Sign}(\cos{(\phi_b - \phi_i)})  \right]} \nn \\
    &\quad\times \prod_{i=1,2}\left( \frac{R^2 b_0^2/(4b_T^2 \cos^2{\phi_b})}{\mu_{ucs}^2} \right)^{-C_i A_{\gamma_{\rm csup}} (\mu_{ucs}, \mu_{f})},
\end{align}
which also contributes to the resummation of logarithms of the jet radius.

To prepare for the NNLL resummation, we now present the renormalized collinear-soft and ultra-collinear-soft functions at NLO accuracy. Their corresponding bare expressions are given in appendix~\ref{app:cs_ucs}. After integrating out the angular dependence, the NLO collinear-soft function contributes to $S_i$ as
\begin{align} \label{eq:cs_1}
    &\langle \bm{1} \otimes \bm{S}^{\mathrm{cs}}_{1}(\{n_i,\bar{n}_i, u_1\}, b_y, R,\mu,\nu) \rangle  \nn\\
    =\,&\frac{\alpha_s(\mu)}{4\pi} C_i \left[ -\ln^2{\frac{\mu^2 b_y^2}{b_0^2}} + 4\ln{\frac{\mu^2 b_y^2}{b_0^2}} \left( \ln \frac{\mu}{\nu} + \ln \frac{2\cosh \eta_i}{R} \right) - \frac{\pi^2}{6} \right] + \mathcal{O}(\alpha_s^2),
\end{align}
where $\bm{S}^{\mathrm{ucs}}_{1} = \bm{1} + \mathcal{O}(\alpha_s)$. Only the tree-level term of $\bm{S}^{\mathrm{ucs}}_{1}$ is required for the NLO expansion of $S_i$ as shown in eq.~\eqref{eq:cs_1}, because $\bm{S}^{\mathrm{cs}}_{m_i} \sim \mathcal{O}(\alpha_s^{m_i})$. 
Meanwhile, the ultra-collinear-soft function contributing to $S_i$ up to NLO accuracy takes the form
\begin{align} \label{eq:ucs_1}
    \langle \bm{S}^{\mathrm{ucs}}_{0}(\{n_i,\bar{n}_i\}, \vec{b}_T, R, \mu) \rangle
    &= 1 + \frac{\alpha_s(\mu)}{4\pi} C_i \left[-4\ln ^2\left(\frac{-2 i \cos \left(\phi_{b}-\phi_i\right) \mu b_T}{b_0 R}\right) - \frac{\pi^2}{2}\right] + \mathcal{O}(\alpha_s^2) \nn\\
    = 1 + \frac{\alpha_s(\mu)}{4\pi} &C_i \left\{-4\left[ \ln \left(\frac{\mu }{b_0 R /(2|b_x|)}\right) -i\frac{\pi}{2}\text{Sign}(n_{i,x} b_x)\right]^2 - \frac{\pi^2}{2}\right\} + \mathcal{O}(\alpha_s^2).
\end{align}
Given that we have $\bm{S}^{\mathrm{cs}}_{0} = \bm{1}$, and no additional collinear parton directions need to be integrated over, we omit the term $\otimes \bm{S}^{\mathrm{cs}}_{0}$ inside the color trace.

In conclusion, the NNLL resummation formula is given by
\begin{align} \label{eq:resum_qT_RRG}
&\,\frac{\mathrm{d}^5 \sigma}{\df\eta_1\, \df\eta_2\, \df p_T\, \df q_T \,\df\phi_q} \nn\\
=\,&  \sum_{ijk\ell} \frac{x_a x_b}{16 \pi \hat{s}^2} \frac{2p_T}{1+\delta_{k\ell}} q_T 
\int_{0}^{2\pi}\!\df\phi_b \int_0^\infty\frac{b_T\, \df b_T}{(2\pi)^2} 
e^{i\vec{q}_T\cdot \vec{b}_T} \notag \\
& \times \sum_{KK'} \exp\left\{ \int_{\mu_h}^{\mu_f} \frac{\mathrm d \mu}{\mu} \left[\gamma_{\text{cusp}}(\alpha_s)  \left( C_H\ln{\frac{\hat{s}}{\mu^2}} + \lambda_K + \lambda_{K'}^* \right) + 2 \gamma_H(\alpha_s) \right]\right\}  \notag \\
&\times\prod_{p=ab}\left(\frac{\nu_{s}} {\nu_{p}}\right)^{\Gamma^{p}_\nu(b_{T},\, \mu_f)}
    \prod_{p=12}\left(\frac{\nu_{cs,p}} {\nu_{p}}\right)^{\Gamma^{p}_\nu(b_y,\, \mu_{cs})} 
    \notag \\
&\times    \prod_{p=12}\exp{\left[\int_{\mu_{cs} }^{\mu_f} \frac{\df\mu}{\mu} \left(\Gamma^{J_p} + \Gamma^{\rm cs}_p \right) 
    +  \int_{\mu_{ucs} }^{\mu_f} \frac{\df\mu}{\mu}  \Gamma^{\rm ucs}_p \right]} 
    U_{\rm NG}^p (\mu_{ucs}, \mu_{cs}) \nn\\
& \times \bm{\mathcal H}_{ij \rightarrow k\ell, KK'}(p_T, \eta_1-\eta_2, \mu_h)  \bm{\mathcal{S}}^{\rm global}_{ijk\ell,K'K}(\vec{b}_T, \eta_1, \eta_2, \mu_f, \nu_{s})  \notag \\
&\times \prod_{p=12}\sum_{m_p=0}^\infty \left\langle \bm{\mathcal{S}}_{m_p}^{\rm ucs}(\{n_p,\bar n_p, \underline{u}\}, \vec{b}_T, R,\mu_{ucs}) \otimes \bm{\mathcal{S}}_{m_p}^{\rm cs}(\{n_p,\bar n_p, \underline{u}\}, b_y, R,\mu_{cs},\nu_{cs,p}) \right\rangle \notag\\
& \times B_{i / p}\left(x_a, b_T, \omega_a,\mu_f, \nu_a \right) B_{j / p}\left(x_b, b_T, \omega_b, \mu_f, \nu_b \right)  \mathcal{J}_k\left(b_y, \omega_1, \mu_{j}, \nu_1 \right) \mathcal{J}_\ell\left(b_y, \omega_2, \mu_{j}, \nu_2 \right) \notag\\
& \times \exp\left[-S^i_{\text{NP}}(b_T,\omega_a,\nu_a) -S^j_{\text{NP}}(b_T,\omega_b,\nu_b) \right] .
\end{align}

To avoid the Landau poles in the resummation formula, we adopt the $b_*$-prescription, 
\begin{align}
    b_{T,*} = \frac{b_T}{\sqrt{1+b_T^2/b_{\rm max}^2}},\quad b_{y,*} = \frac{|b_y|}{\sqrt{1+b_y^2/b_{\rm max}^2}},\quad b_{x,*} = \frac{|b_x|}{\sqrt{1+b_x^2/b_{\rm max}^2}}.
\end{align}
With the exception of the hard scale $\mu_h = v_h\, 2p_T$, all other perturbative scales are obtained through this prescription as
\begin{align}
     \mu_s = \mu_b = \mu_f = \min{\left\{v_f \frac{b_0}{b_{T,*}}, \,\mu_h\right\}}, \quad \mu_j = \mu_{cs} = v_f\frac{b_0}{b_{y,*}}, \quad \mu_{ucs} = \min{\left\{v_f \frac{Rb_0}{2b_{x,*}}, \,\mu_{cs}\right\}}.
\end{align}
With $v_h$ and $v_f$ varied in the range $[0.5,\,2]$, we evaluate the theoretical uncertainty associated with the hard scale, as well as the uncertainties from $\mu_f$, $\mu_{cs}$, and $\mu_{ucs}$, by varying them simultaneously, as discussed in section~\ref{sec:numerical}. To respect the RG flow from the ultraviolet to the infrared, we require the factorization scale $\mu_f$ to be bounded by the hard scale $\mu_h$, $\mu_f \leq \mu_h$. Furthermore, to preserve the scale hierarchy inherent to the small-$R$ refactorization and ensure the positivity of the evolution variable $u$ in eq.~\eqref{eq:U_NGL}, we impose the constraint $\mu_{ucs} \leq \mu_{cs}$.

\section{Numerical results} \label{sec:numerical}

In this section, we present our numerical predictions for the azimuthal decorrelation $\delta\phi$ and the transverse momentum imbalance $q_T$ in inclusive dijet production at the LHC. We perform the calculation at a center-of-mass energy of $\sqrt{s} = 13$ TeV. Jets are identified using the anti-$k_T$ clustering algorithm~\cite{Cacciari:2008gp} with a radius parameter $R=0.5$ and the standard $E$-scheme~\cite{Cacciari:2011ma} for recombination. 
The kinematic acceptance is defined by the following cuts on the standard $E$-scheme transverse momenta of the leading ($p_{T,1}$) and subleading ($p_{T,2}$) jets, as well as their rapidities,
\begin{equation} \label{eq:jet_cuts}
    p_{T,1} > 100~\mathrm{GeV} , \quad 
    p_{T,2} > 80~\mathrm{GeV} , \quad 
    |\eta_{1,2}| < 2.0 \,.
\end{equation}
As detailed in the previous sections, the observables $\delta\phi$ and $q_T$ are defined with respect to the reconstructed winner-take-all (WTA) axis\footnote{To avoid losing particles due to the mismatch between the WTA axis and the original axis, we recluster in the WTA scheme with a larger jet radius.}, which minimizes sensitivity to NGLs.

For the fixed-order components required for the NLO matching procedure, we utilize \texttt{NLOJET++} and \texttt{FASTJET} to compute the LO differential cross sections for the 3-parton production process. We also compare our matched predictions with parton-shower results obtained from \texttt{PYTHIA}~8 to assess the impact of non-perturbative effects.

\subsection{NNLL resummation}
\begin{figure}[t]
    \centering
    \begin{subfigure}[t]{0.49\textwidth}
        \centering
        \includegraphics[width=\linewidth]{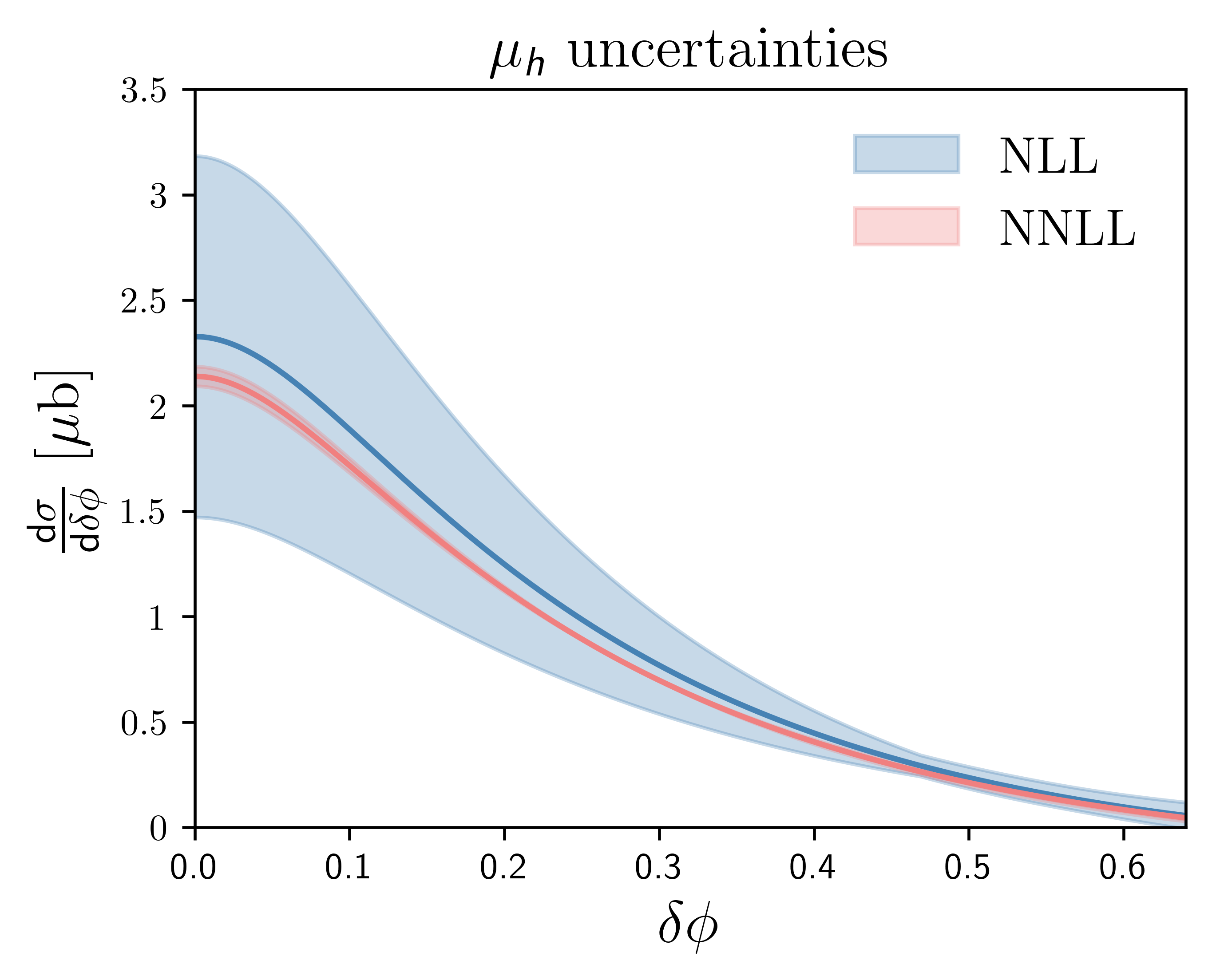}
        \caption{$\delta\phi$ distribution}
        \label{fig:dphi_muh_uncer}
    \end{subfigure}
    \hfill
    \begin{subfigure}[t]{0.49\textwidth}
        \centering
        \includegraphics[width=\linewidth]{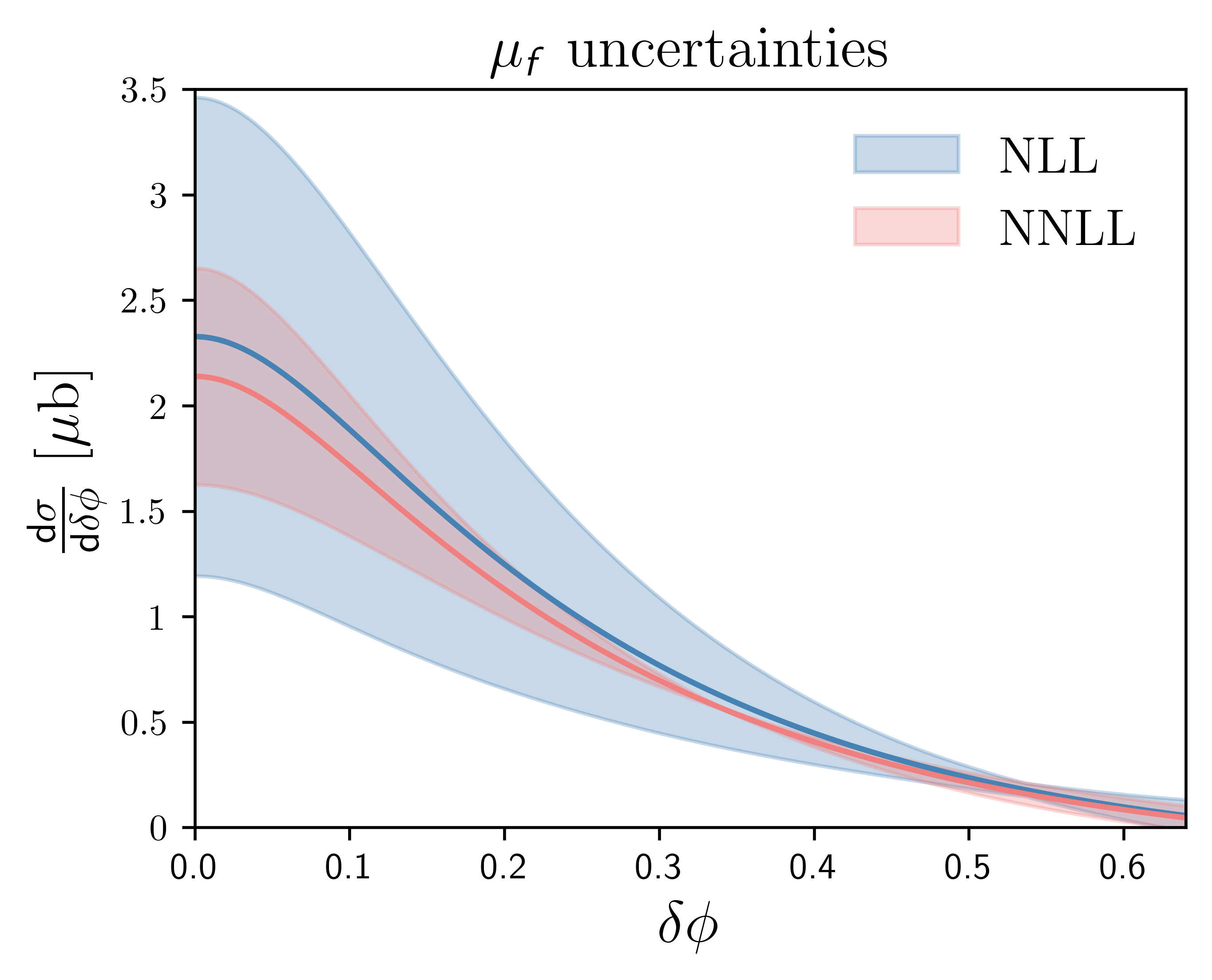}
        \caption{$\delta\phi$ distribution}
        \label{fig:dphi_mubs_uncer}
    \end{subfigure}
    
    \vspace{0.3cm}
    
    \begin{subfigure}[t]{0.49\textwidth}
        \centering
        \includegraphics[width=\linewidth]{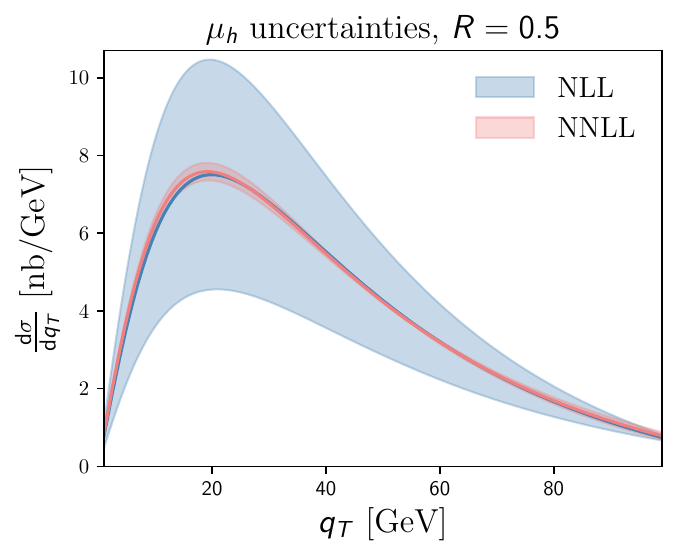}
        \caption{$q_T$ distribution}
        \label{fig:qT_muh_uncer2}
    \end{subfigure}
    \hfill
    \begin{subfigure}[t]{0.49\textwidth}
        \centering
        \includegraphics[width=\linewidth]{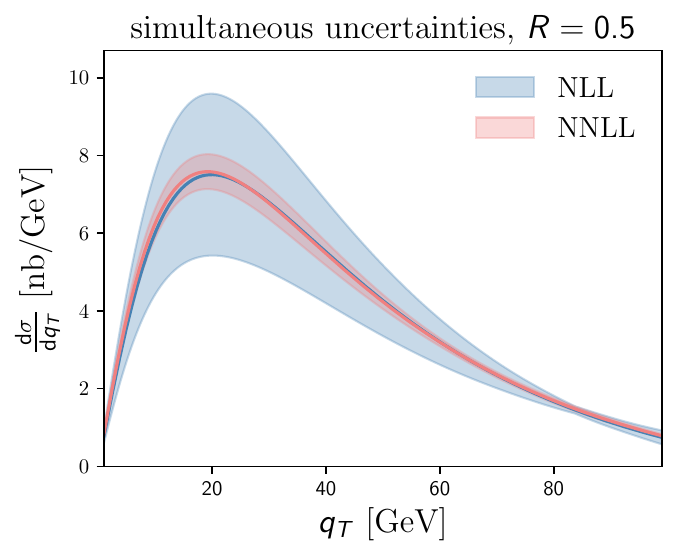}
        \caption{$q_T$ distribution}
        \label{fig:qT_muf_uncer2}
    \end{subfigure}
    
    \caption{Theoretical uncertainties for the resummed $\delta\phi$ (top row) and $q_T$ (bottom row, $R=0.5$) distributions. The blue and red bands represent NLL and NNLL predictions, respectively. The left column (a, c) shows the uncertainty derived from varying the hard scale $\mu_h$ by factors of 2 and 0.5 around its central value $2p_T$. The right column (b, d) displays the scale uncertainties from the factorization scale $\mu_f$ (for $\delta\phi$) and the simultaneous variation of factorization, collinear-soft and ultra-collinear-soft scales (for $q_T$).}
    \label{fig:resum_uncertainty}
\end{figure}

In figure~\ref{fig:resum_uncertainty}, we present the resummed differential cross sections for the azimuthal decorrelation $\delta\phi$ (top row) and the transverse momentum imbalance $q_T$ (bottom row). The blue bands correspond to NLL accuracy, while the red bands denote the NNLL results. We observe a substantial reduction in theoretical scale uncertainties for both observables when advancing from NLL to NNLL accuracy.

For the $\delta\phi$ distribution, as shown in figure~\ref{fig:resum_uncertainty}(a), the uncertainty associated with the hard scale $\mu_h$ becomes almost negligible at NNLL. The factorization scale uncertainty (figure~\ref{fig:resum_uncertainty}(b)) also decreases markedly, though it remains the dominant source of theoretical uncertainty. Crucially, the NNLL bands lie consistently within the NLL envelopes, demonstrating the excellent perturbative convergence of our factorization framework.

Similar behavior is observed for the $q_T$ distribution, calculated with $R=0.5$. Figure~\ref{fig:resum_uncertainty}(c) shows a drastic reduction in hard scale dependence at NNLL. Figure~\ref{fig:resum_uncertainty}(d) presents the uncertainty arising from the simultaneous variation of the factorization scale $\mu_f$ and the associated soft scales ($\mu_{cs}$ and $\mu_{ucs}$). While this remains the dominant uncertainty, the bandwidth is substantially reduced. The distribution peaks in the perturbative region around $q_T \sim 18$--$22$ GeV, and the overlap between the NLL and NNLL orders confirms the reliability of the resummation framework.

\begin{figure}[t]
    \centering
    \includegraphics[width=0.6\linewidth]{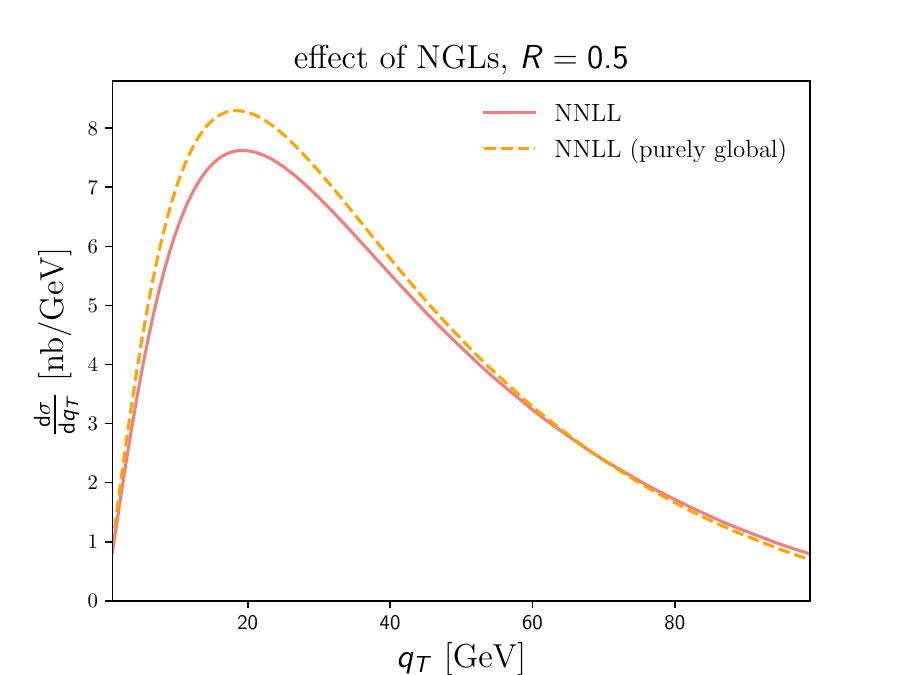}
    \caption{The impact of the resummation of NGLs on the $q_T$ distribution for a jet radius of $R=0.5$. The solid red curve represents the full NNLL result including the non-global resummation factor $U_{\mathrm{NG}}$, while the dashed orange curve shows the ``purely global'' NNLL prediction where the contribution from $U_{\mathrm{NG}}$ is turned off.}
    \label{fig:UNG_effect}
\end{figure}

We have also investigated the effect on the $q_T$ distribution of the resummation of NGLs by $U_{\rm NG}$, shown in figure~\ref{fig:UNG_effect}.

\subsection{Power corrections and NLO matching}

\subsubsection{Power corrections}

\begin{figure}[t]
    \centering
    \begin{subfigure}[b]{\linewidth}
        \includegraphics[width=\linewidth]{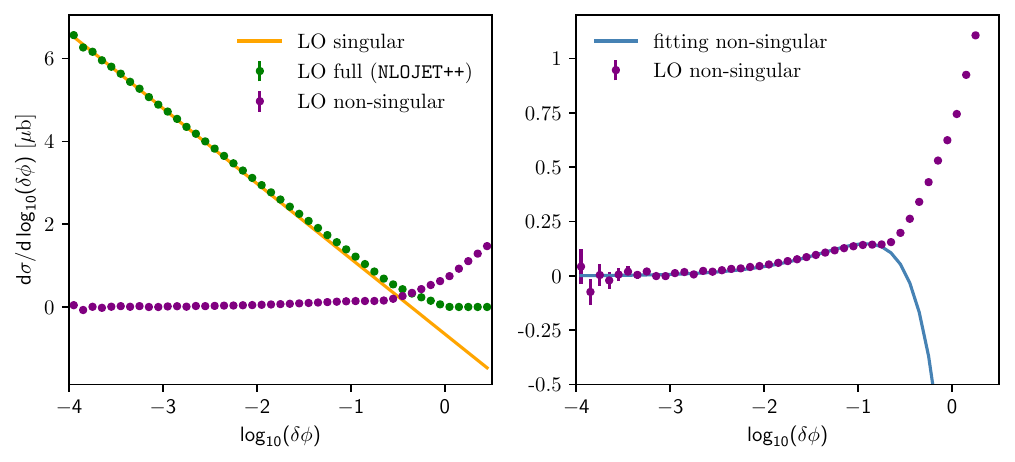}
        \caption{The azimuthal decorrelation $\delta\phi$.}
        \label{fig:power_dphi}
    \end{subfigure}
    
    \vspace{0.5cm} 
    
    \begin{subfigure}[b]{\linewidth}
        \includegraphics[width=\linewidth]{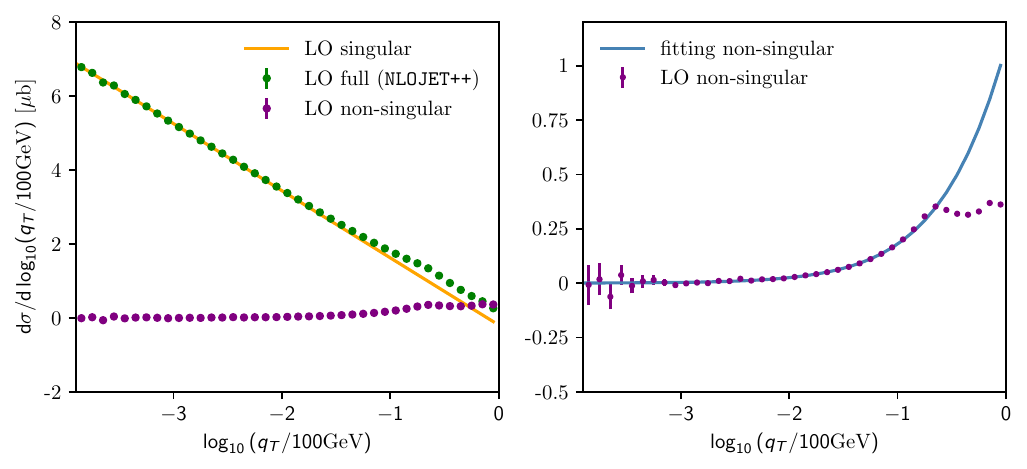}
        \caption{The transverse momentum imbalance $q_T$.}
        \label{fig:power_qT}
    \end{subfigure}
    
    \caption{
    Validation of factorization and power corrections for the observables (a) and (b). In the left column, the full LO differential cross sections from \texttt{NLOJET++} (green points) are compared with the singular contributions predicted by our factorization formulae (solid orange lines); their difference yields the non-singular terms (purple points). In the right column, the non-singular power corrections are fitted to the functional form $f(\mathcal{O}) = a \mathcal{O} + b \mathcal{O} \ln \mathcal{O}$ (blue lines). The fit is performed over the range $\log_{10}(\delta\phi)\in[-3.0, -0.7]$ for azimuthal decorrelation, and $\log_{10}(q_T/100\,\text{GeV}) \in [-3.9, -1.0]$ for transverse momentum imbalance. The kinematic cuts imposed on the jets are summarized in eq.~\eqref{eq:jet_cuts}.}
    \label{fig:power_corrections}
\end{figure}

In this section, we validate the singular components of our factorization formulae against fixed-order results generated by \texttt{NLOJET++} at LO, representing the Born-level 3-parton production. Figure~\ref{fig:power_corrections} displays the results for the azimuthal decorrelation $\delta\phi$ (top row) and the transverse momentum imbalance $q_T$ (bottom row).

The left panels of figure~\ref{fig:power_corrections} demonstrate that in the deep infrared regions, specifically $\delta\phi \ll 1$ and $q_T \ll p_T$, the predicted singular distributions (solid orange lines) converge precisely with the full LO calculations (green points). This agreement confirms the accuracy of the singular factorization formulae derived in sections~\ref{sec:dphi_fact} and \ref{sec:qT_fact}.

In the right panels, we isolate the non-singular power corrections (purple points), defined as the difference between the full theory and the singular terms. To capture the behavior of these corrections, we fit the data (blue curves) using a logarithmic ansatz of the form $a \mathcal{O} + b \mathcal{O} \ln \mathcal{O}$, where $\mathcal{O}$ represents either $\delta\phi$ or $q_T$. For instance, the $\delta\phi$ corrections are well described by this function over the range $\log_{10}(\delta\phi) \in [-3.0, -0.7]$.

To interpret the physical implications of this fit, we decompose the cross section into singular and power-correction terms. The logarithmic distribution (as plotted) and the linear differential distribution are given by
\begin{align} \label{eq:lin_dist}
    \frac{\mathrm{d}\sigma ({\rm LO \,\, full})}{\mathrm{d} \log_{10}\mO} &\simeq \frac{\mathrm{d}\sigma ({\rm LO \,\, singular})}{\mathrm{d} \log_{10}\mO} + \frac{\mathrm{d}\sigma ({\rm power\, corrections})}{\mathrm{d} \log_{10}\mO} \nn\\
    &= (A_1 \ln{\mO} + A_0) + \left( B_1\mO + B_0\mO\ln{\mO}\right), \nn\\
    \frac{\mathrm{d}\sigma ({\rm LO \,\, full})}{\mathrm{d}\mO}
    &\simeq  \frac{A_1\ln{\mO} + A_0}{\mO} + \left(B_1 + B_0\ln{\mO}\right).
\end{align}
Here, the coefficients $A_1$ and $A_0$ parameterize the leading-power singular behavior, while $B_1$ and $B_0$ characterize the non-singular contributions. Notably, as shown in eq.~\eqref{eq:lin_dist}, the non-singular contribution $(B_1 + B_0\ln\mathcal{O})$ retains a logarithmic divergence as $\mathcal{O} \to 0$, despite being suppressed relative to the leading singular terms. 

This logarithmic enhancement is characteristic of exclusive or jet-dependent observables \cite{Ebert:2019zkb, Salam:2021tbm, Grazzini:2017mhc, Alekhin:2021xcu, Chen:2026zmi}. This stands in contrast to the behavior of inclusive quantities, such as the Drell-Yan transverse momentum spectrum, which takes the form (see e.g.,~\cite{Ebert:2018gsn, Becher:2020ugp}) 
\begin{equation}
\frac{\mathrm{d}\sigma^\mathrm{LO}}{\mathrm{d} q_T} = \frac{A_1\ln q_T + A_0}{q_T} + (C_1 \ln q_T + C_0)q_T +\dots,
\end{equation}
where $A_i$ and $C_i$ are constant coefficients. In that framework, the power corrections ($(C_1 \ln q_T + C_0)q_T$) vanish as $q_T \to 0$. Conversely, the presence of the $b \ln \mathcal{O}$ term in our analysis implies that power corrections in our paper are more singular than those in inclusive Drell-Yan, diverging logarithmically rather than vanishing in the infrared limit.

\subsubsection{NLO matching}
The explicit expression for the \emph{naive} additive matching prescription, used to match the NNLL resummation curve to the LO fixed-order result, is
\begin{equation}
    \df \sigma_{\rm add} ({\rm NNLL}+{\rm LO}) = \df\sigma({\rm NNLL}) + \underbrace{\df\sigma({\rm LO \,\, full}) - \df\sigma({\rm LO \,\,singular})}_{\df\sigma({\rm LO \,\,non-singular})}.
\end{equation}
The $\df\sigma(\text{LO non-singular})$ term defined here, also known as the matching correction, is dominated by $\df\sigma ({\rm power\, corrections})$ in the small $\mO$ region.

However, two issues remain to be resolved in the matching procedure. 
First, in the small-$\mO$ region, divergent power corrections of the form $a + b\ln\mO$ degrade the quality of the matching, which is the main reason for the poor performance of the naive additive scheme in our case. Second, since factorization breaks down in the large-$\mO$ region and the resummed result contains Sudakov effects beyond the fixed-order accuracy, this unphysical contribution must be removed to recover the pure fixed-order prediction.

\begin{figure}[t]
    \centering
    \begin{minipage}{1\linewidth}
    \begin{subfigure}{0.49\textwidth}
        \includegraphics[width=\linewidth]{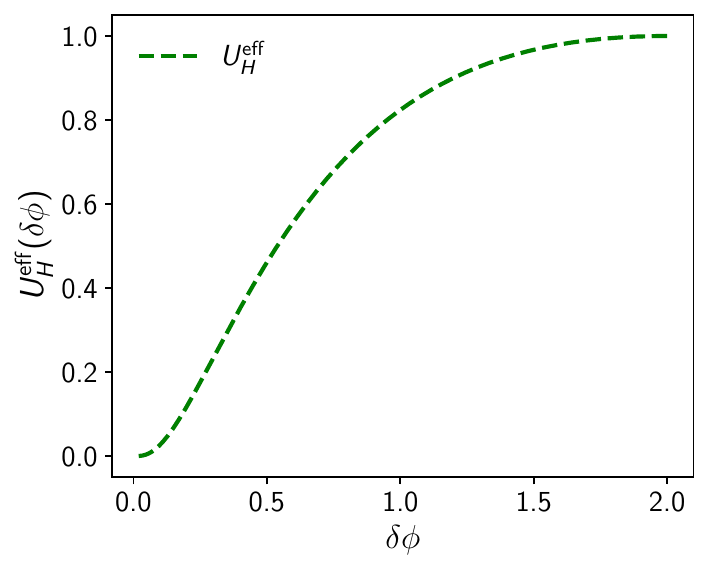}
    \end{subfigure}
    \begin{subfigure}{0.49\textwidth}
        \includegraphics[width=\linewidth]{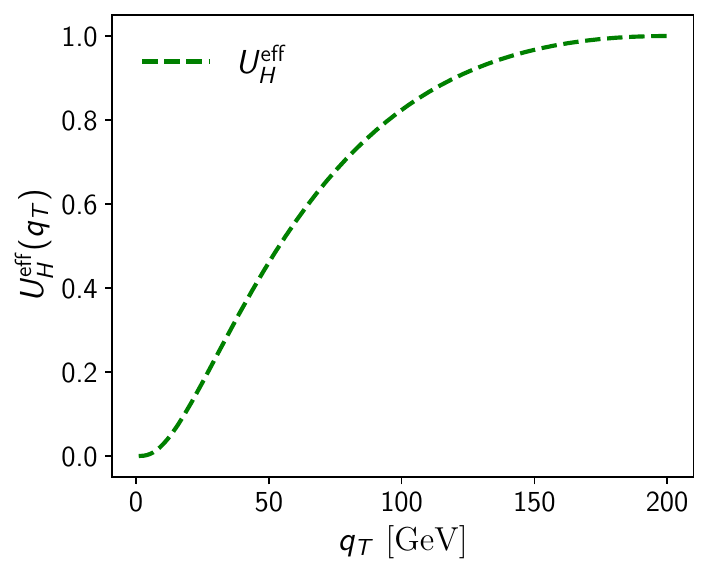}
    \end{subfigure}
    \end{minipage}
    \begin{minipage}{1\linewidth}
    \begin{subfigure}{0.49\textwidth}
        \includegraphics[width=\linewidth]{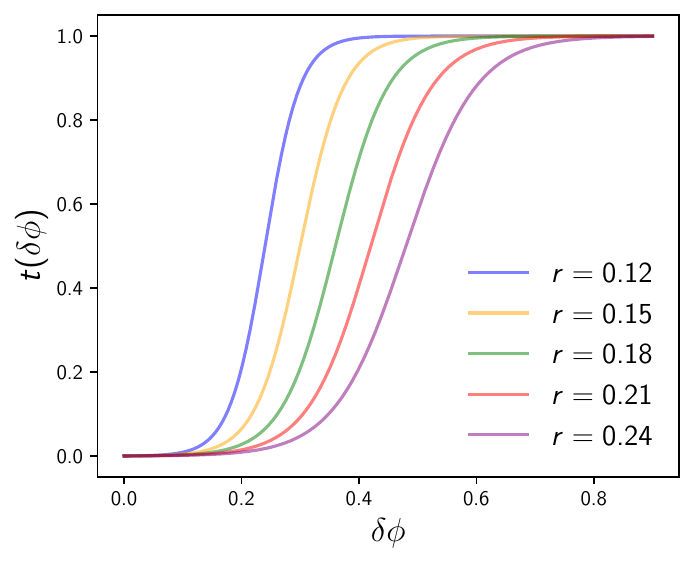}
    \end{subfigure}
    \begin{subfigure}{0.49\textwidth}
        \includegraphics[width=\linewidth]{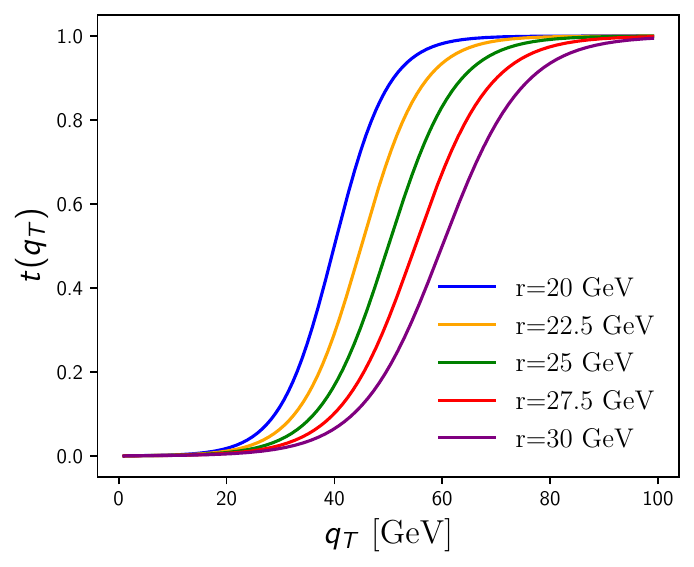}
    \end{subfigure}
    \end{minipage}
    
    \caption{Key components of the modified additive matching scheme. The top row displays the effective evolution matrix $U_H^{\rm eff}(\mathcal{O})$ for $\mathcal{O}=\delta\phi$ (left) and $q_T$ (right). This factor is applied to the matching correction to suppress the divergence of power corrections in the deep infrared region ($\mathcal{O} \to 0$). The bottom row shows the transition profiles $t(\delta\phi)$ (left) and $t(q_T)$ (right) for various transition parameters $r$. These functions control the smooth switch from the resummed prediction to the fixed-order result at large kinematic values.}
\label{fig:trans_func}
\end{figure}

To address the first issue, we utilize an effective evolution matrix $U_H^{\rm eff}$ to resum the large logarithm $\ln{\mO}$ in the small $\mO$ region of matching corrections,
\begin{align} \label{eq:UHeff}
    U_H^{\rm eff}(\mO)& = \exp{ \left[ \int_{2p_T}^{\mu_\mO} \frac{\df\mu}{\mu} \left( \frac{\alpha_s (2p_T)}{\pi} 4C_A \ln{\frac{4p_T^2}{\mu^2}} \right) \right]}, 
\end{align}
with $p_T = 100~\text{GeV}$.
For the $\delta\phi$ distribution, we set $\mu_\mO = p_T \delta\phi$, while $\mu_\mO = q_T$ for the $q_T$ distribution. 
Eq.~\eqref{eq:UHeff} corresponds to the evolution matrix of the $gg\to gg$ channel at double-logarithmic accuracy, with the coupling constant fixed at the hard scale $2p_T$ and non-diagonal anomalous dimensions neglected. 
The effective Sudakov factors for $q_T$ and $\delta\phi$ distribution are shown in figure~\ref{fig:trans_func}.

We have explored the impact of using a running coupling scheme in eq.~\eqref{eq:UHeff}, which would formally probe some NLL effects, but found that this would only lead to small changes to the effective Sudakov factor.

As for the second issue, a transition function $t(\mO)$ can be introduced to smoothly switch off the resummation and transition to the fixed order:
\begin{align}
    t(\mO) = \frac{1}{2} - \frac{1}{2}\tanh{\left(4 - \frac{2\mO}{r}\right)},\quad \mathcal{O} \in \{q_T, \delta\phi\} .
\end{align}

Accordingly, the modified additive matching scheme is defined as 
\begin{align}
    \df \sigma_{\rm match} ({\rm NNLL}+{\rm LO}) &= (1-t(\mO))\left(\df\sigma({\rm NNLL}) + U_{H}^{\rm eff}(\mO) ~ \df\sigma({\text{LO non-singular}}) \right) \nn\\
    &\quad + t(\mO)~ \df\sigma({\rm LO \,\, full}).
\end{align}

Figure~\ref{fig:trans_func} illustrates the behavior of the two regulating functions introduced in our modified matching scheme. The top panels depict the effective evolution matrix, $U_H^{\rm eff}(\mathcal{O})$. As the observable $\mathcal{O}$ approaches zero, the Sudakov suppression causes $U_H^{\rm eff}$ to vanish. This behavior is crucial: it effectively dampens the divergent non-singular contributions (discussed in the previous section) in the deep infrared limit, thereby stabilizing the numerical matching where the naive additive scheme would fail.

The bottom panels of figure~\ref{fig:trans_func} display the transition function $t(\mathcal{O})$ for different values of the profile parameter $r$. This function acts as a smooth switch, transitioning from $0$ in the resummation-dominated region to $1$ in the fixed-order region. By varying $r$, we shift the transition point, allowing us to estimate the uncertainty associated with the matching procedure. Concretely, for matching $\delta\phi$, we vary $r\in[0.12, 0.24]$, while for $q_T$, we vary $r\in[20,30]\text{~GeV}$.
This mechanism ensures that unphysical Sudakov effects are removed at large $\delta\phi$ or $q_T$, seamlessly recovering the correct LO cross section in the tail of the distribution.

\begin{figure}[t]
    \centering
    \begin{subfigure}{0.49\textwidth}
        \includegraphics[width=\linewidth]{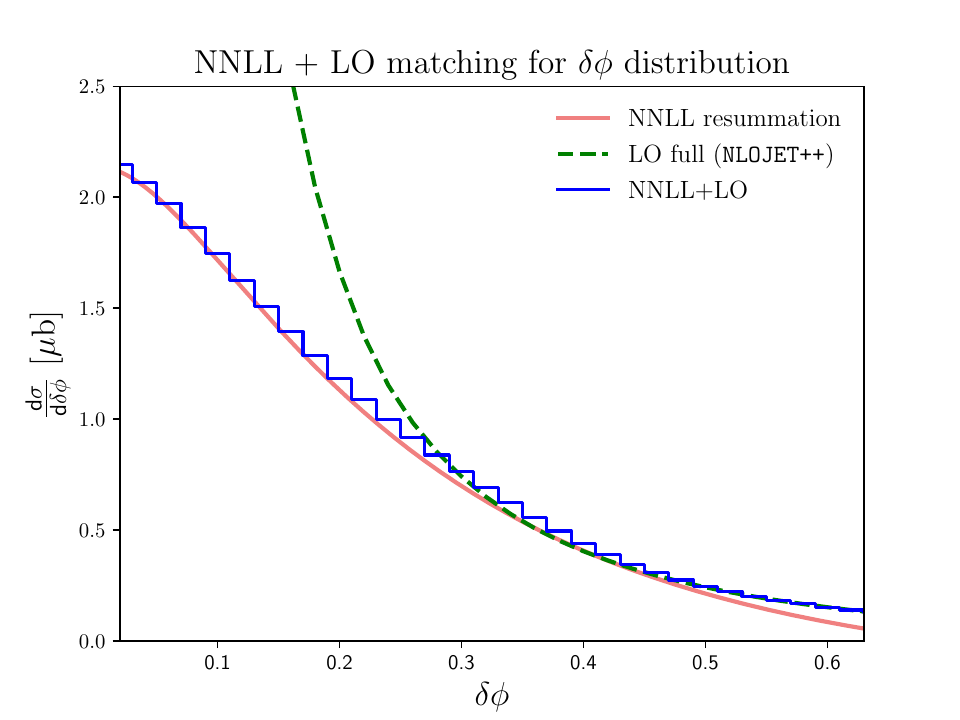}
        \caption{}
        \label{fig:match_dphi_abs}
    \end{subfigure}
    \hfill
    \begin{subfigure}{0.49\textwidth}
        \includegraphics[width=\linewidth]{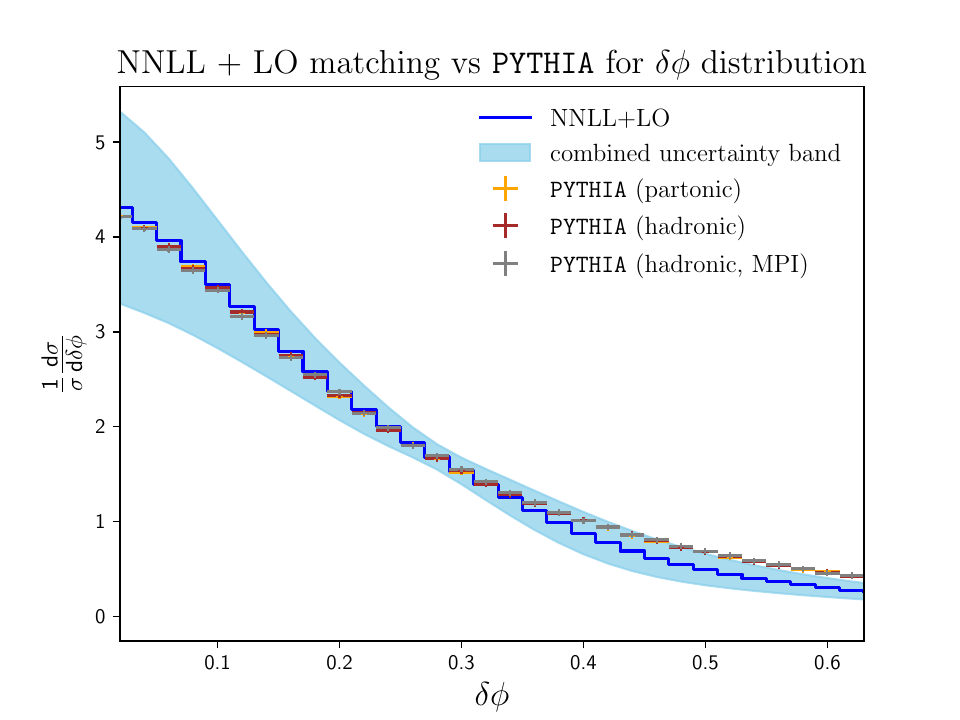}
        \caption{}
        \label{fig:match_dphi_pythia}
    \end{subfigure}
    
    \vspace{0.5cm} 
    
    \begin{subfigure}{0.49\textwidth}
        \includegraphics[width=\linewidth]{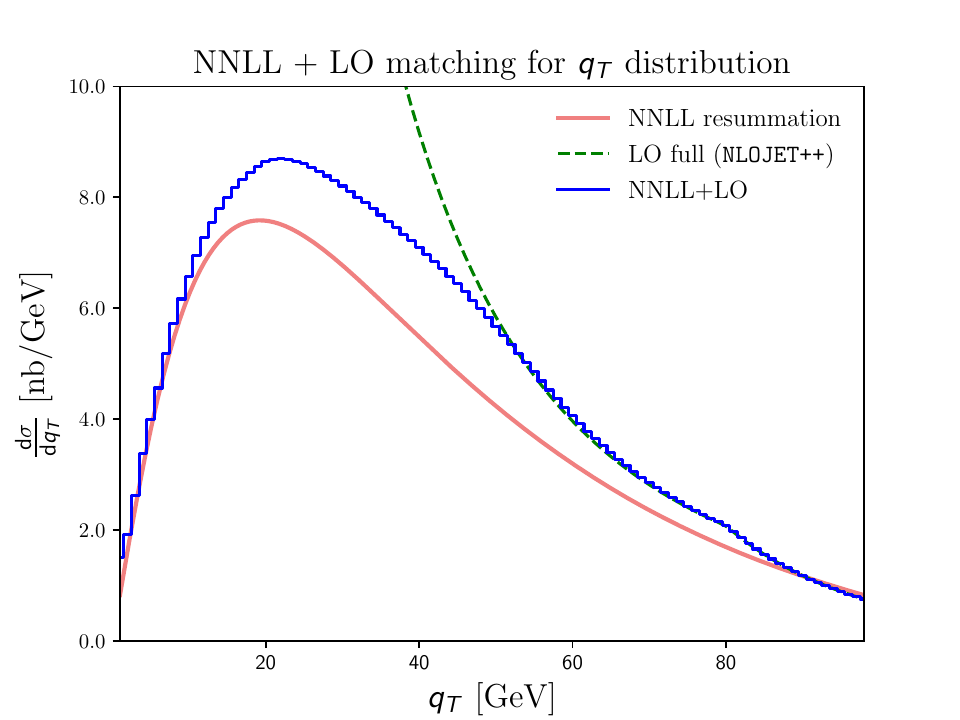}
        \caption{}
        \label{fig:match_qT_abs}
    \end{subfigure}
    \hfill
    \begin{subfigure}{0.49\textwidth}
        \includegraphics[width=\linewidth]{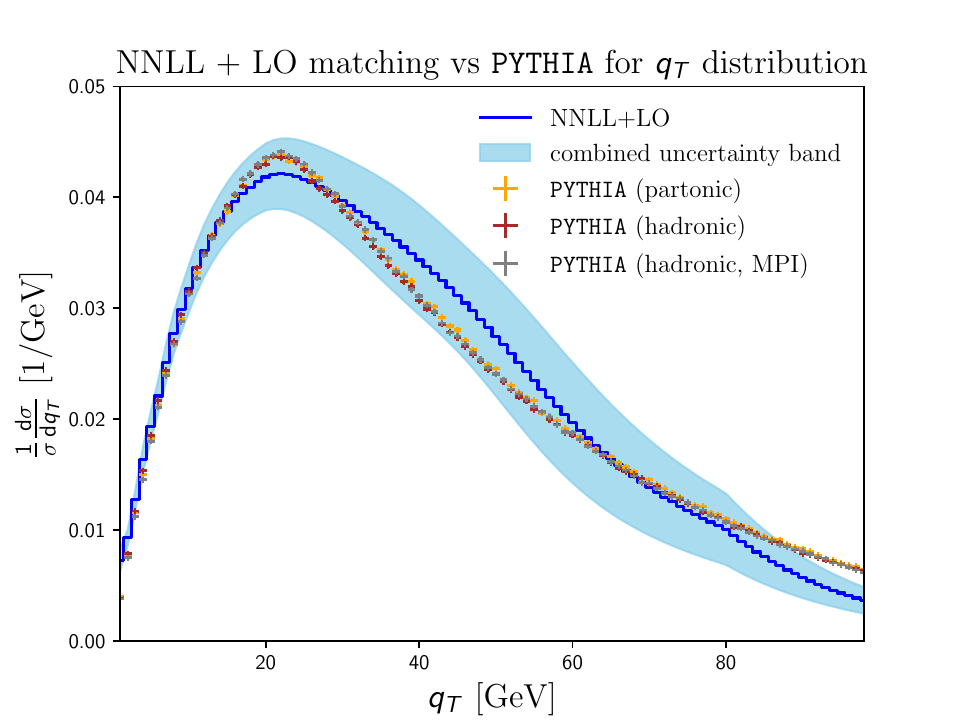}
        \caption{}
        \label{fig:match_qT_pythia}
    \end{subfigure}
    
    \caption{Matched distributions and comparison to \texttt{PYTHIA}.
    The kinematic cuts imposed on the jets are summarized in eq.~\eqref{eq:jet_cuts}.
    The top row displays the azimuthal decorrelation $\delta\phi$ (using transition parameter $r=0.18$), while the bottom row displays the transverse momentum imbalance $q_T$ ($r=25$ GeV). The left column (a, c) shows the absolute differential cross sections. The right column (b, d) compares the shapes to \texttt{PYTHIA} results, with curves normalized to unity in the ranges $\delta\phi\in[0,0.45]$ and $q_T\in[0,30]~\text{GeV}$, respectively. The total uncertainty bands include contributions from resummation scale variations, fixed-order scale variations, and the transition parameter $r$, added in quadrature.}
    \label{fig:matching_combined}
\end{figure}

The final matched results for the $\delta\phi$ and $q_T$ distributions are presented in figure~\ref{fig:matching_combined}, with the top and bottom rows corresponding to $\delta\phi$ and $q_T$, respectively. In the left panels (figures~\ref{fig:matching_combined}(a) and \ref{fig:matching_combined}(c)), we show the absolute cross sections obtained using the modified additive matching scheme. 

In the right panels (figures~\ref{fig:matching_combined}(b) and (d)), we compare the normalized shapes of our parton-level analytical predictions (NNLL matched to LO) with results from \texttt{PYTHIA}~8. 
The theoretical uncertainty bands incorporate the resummation and fixed-order scale variations, along with the sensitivity to the transition parameter $r$, combined in quadrature. These bands illustrate the stability of the matching procedure across the transition from the resummation-dominated region to the fixed-order tail.
To evaluate non-perturbative effects, the \texttt{PYTHIA}~8 results are shown at the parton level (yellow), hadron level (red), and hadron level with multi-parton interactions (MPI) (grey).

For the azimuthal decorrelation $\delta\phi$, the NNLL+LO calculation agrees well with the Monte Carlo results at small $\delta\phi$. At large $\delta \phi$, the observed deviations are expected and primarily driven by the absence of matching in \texttt{PYTHIA}. Furthermore, the three \texttt{PYTHIA}~8 predictions are nearly identical, indicating that $\delta\phi$ is largely insensitive to hadronization and MPI in this kinematic regime.

A similar behavior is observed for the $q_T$ distribution (figure~\ref{fig:matching_combined}(d)), where the analytic predictions are consistent with \texttt{PYTHIA}~8, and the Monte Carlo results fall well within the theoretical uncertainty bands. As with $\delta\phi$, the non-perturbative corrections in \texttt{PYTHIA}~8 remain negligible. This confirms that for the chosen jet radius and kinematic cuts, the distribution is dominated by perturbative physics and is accurately captured by our resummation framework.

\section{Conclusions}
\label{sec:conclusions}

In this work, we have presented a comprehensive study of the azimuthal decorrelation $\delta\phi$ and the transverse momentum imbalance $q_T$ in inclusive dijet production at the LHC. A central feature of our analysis is the use of the winner-take-all (WTA) recombination scheme for defining the jet axes. By adopting the WTA scheme, we eliminate the NGLs that plague standard jet definitions, thereby simplifying the all-order structure of the cross section.

We derived the factorization formulae for both observables using SCET. We demonstrated that, despite the common goal of measuring back-to-back kinematics, $\delta\phi$ and $q_T$ exhibit distinct factorization structures. A particularly intricate aspect of our analysis was the treatment of the $q_T$ distribution in the small jet radius limit ($R \ll 1$). We showed that the soft function requires a further refactorization into global soft, collinear-soft, and ultra-collinear-soft modes. Although the WTA scheme removes the standard NGLs of $q_T$, there are NGLs of $R$. Correspondingly, the scale hierarchy between collinear-soft and ultra-collinear-soft modes yields the non-trivial NGL evolution structure.

Numerically, we matched our resummed predictions to fixed-order LO calculations. For the azimuthal decorrelation $\delta\phi$, we achieved full NNLL accuracy. For the transverse momentum imbalance $q_T$, we implemented a joint resummation involving NNLL accuracy for global logarithms and LL accuracy for the small-$R$ NGLs. Compared to NLL, our higher-order accuracy results in a substantial reduction of theoretical uncertainties. Our numerical analysis confirms that the scale-variation bands at NNLL (combined with LL accuracy for NGLs for $q_T$) lie consistently within the lower-order envelopes, indicating the robust convergence of our perturbative framework.

Finally, studies with \texttt{PYTHIA}~8 show that both the azimuthal decorrelation $\delta\phi$ and the transverse momentum imbalance $q_T$ are robust against hadronization and multi-parton interactions. The agreement with our predictions is good; the primary discrepancies likely arise from the fixed-order matching, which is included in our calculations but absent in \texttt{PYTHIA}.

Our formalism provides a precise theoretical tool for probing the dynamics of QCD radiation in jet production at the LHC. The techniques developed here, particularly the handling of small-$R$ refactorization and logarithmic power corrections, are applicable to a wide class of jet substructure observables and can facilitate future extractions of transverse-momentum-dependent (TMD) distributions.

\acknowledgments
R.J.F. and D.Y.S. are supported by the National Science Foundations of China under Grant No.~12275052, No.~12147101, No.~12547102. D.Y.S. is also supported by the Innovation Program for Quantum Science and Technology under grant No. 2024ZD0300101. R.R.~is supported by the European Union’s Horizon Europe research and innovation programme under the Marie Sk\l{}odowska-Curie project ``SoftSERVE-NGL'' with grant agreement No. 101108359. B.W.~is supported by the Ram\'{o}n y Cajal program with the Grant No. RYC2021-032271-I, the Xunta de Galicia under the ED431F 2023/10 project, the European Research Council project ERC-2018-ADG-835105 YoctoLHC, the project Mar\'{\i}a de Maeztu CEX2023-001318-M financed by MCIN/AEI/\-10.13039/\-501100011033, by European Union ERDF, and by the Spanish Research State Agency under project PID2020-119632GBI00 and PID2023-152762NB-I00.

\appendix

\section{NLO global soft functions} \label{app:soft_global}
We choose a reference frame in which the $zx$-plane coincides with the scattering plane, and the beam ($i = a,b$) and jet ($i = 1,2$) directions are specified by
\begin{align} 
    n_a^\mu = (1,0,0,1)&,\quad n_b^\mu = (1,0,0,-1)
    ,\quad \nn\\
    n_1^\mu = (1,\sech{\eta_1},0,\tanh{\eta_1})&,\quad n_2^\mu = (1,-\sech{\eta_2},0,\tanh{\eta_2}). \nn
\end{align}
Since at Born level the two jets are produced back-to-back in the transverse plane, their azimuthal angles can be fixed to $\phi_1=0$ and $\phi_2=\pi$. The measurement 4-vector for $q_T$ corresponds to  $b^\mu = (0,\vec{b}_T, 0) = (0, b_T\cos{\phi_b}, b_T\sin{\phi_b}, 0)$ in impact-parameter space.

As discussed in section~\ref{sec:soft_refact}, the $q_T$ soft function can be refactorized into global soft, collinear-soft, and ultra-collinear-soft contributions. In this appendix we present the explicit expression for the bare global-soft function at one loop level,
\begin{align} \label{eq:bare_global}
     \bm S^{\mathrm{global}}_{ijk\ell}\bigl(\vec{b}_T, \eta_1,\eta_2, \epsilon,\eta\bigr) 
    &= \sum_{i<j}\bigl(-\mathbf{T}_i\cdot\mathbf{T}_j\bigr) ~ \, \omega^{\text {global}}_{ij}(\vec{b}_T, \eta_1, \eta_2, \epsilon, \eta) + \mathcal{O}(\alpha_s^2),
\end{align}
where $\epsilon$ comes from dimensional regularization, and $\eta$ is the rapidity regulator. Throughout this work, we use the $\overline{\mathrm{MS}}$ renormalization scheme,
\begin{align}
    g_{s,0}^2 = 4\pi\left(\frac{\mu^2e^{\gamma_E}}{4\pi}\right)^{\!\epsilon} \alpha_s(\mu) Z_{\alpha}, \quad
    Z_{\alpha} = 1 - \frac{\alpha_s(\mu)}{4\pi} \frac{\beta_0}{\epsilon} + \mathcal{O}\left( \alpha_s^2 \right).
\end{align}

The NLO dipole contributing to the soft function can be written as
\begin{align}
     \omega^{\text {global }}_{ij}(\vec{b}_T, \eta_1, \eta_2, \epsilon, \eta) =  
     \frac{\alpha_s(\mu) \mu^{2\epsilon}\pi^{\epsilon}e^{\gamma_E\epsilon}}{\pi^2} 
    \int \mathrm{d}^d k\, \delta(k^2)\theta(k^0)\, \omega^2 \left(\frac{\nu}{2k^0}\right)^\eta\frac{n_i\cdot n_j}{n_i\cdot k ~ n_j\cdot k} e^{i \vec{k}_T\cdot\vec{b}_T}\,,
\end{align}
where the factor $[\nu/(2k^0)]^\eta$ regulates the rapidity divergence that arises when the dipole involves an initial-state leg, i.e., for the beam–beam and beam–jet dipoles. The auxiliary bookkeeping parameter $\omega$, introduced alongside the rapidity regulator, satisfies \cite{Lubbert:2016rku}
\begin{align}
    \frac{\df }{\df \nu} \omega = -\frac{\eta}{2} \omega, \quad \lim_{\eta\to0}\omega\to 1,
\end{align}
and ensures that the bare soft functions remain independent of the rapidity scale, in close analogy to the role of the renormalized coupling constant. 
After first expanding around $\eta \to 0$ and subsequently around $\epsilon \to 0$, the NLO beam-beam soft dipole is given by
\begin{align}
    \omega^{\text {global }}_{ab}(\vec{b}_T, \epsilon, \eta) = \frac{\alpha_s(\mu)}{4\pi}\left[\frac{4}{\epsilon^2} +
    4\left(\ln\frac{\mu^2}{\nu^2} - \frac{2}{\eta}\right) \left(\frac{1}{\epsilon} + \ln\frac{\mu^2b_T^2}{b_0^2}\right)
    - 2\ln^2\frac{\mu^2b_T^2}{b_0^2} -\frac{\pi^2}{3} \right].
\end{align}
The beam-jet dipole, with $i\in\{a,b\}$ and $j\in\{1,2\}$, takes the form
\begin{align}
    \omega^{\text {global }}_{ij}(\vec{b}_T, \eta_j, \epsilon, \eta)= \frac{\alpha_s(\mu)}{4\pi}&\left\{4\left[ -\frac{1}{\eta}+\ln \frac{\mu}{\nu} + \text{Sign}(n_{i,x}) \eta_j + \ln \left(-2 i \cos \left(\phi_{b}-\phi_j\right)\right) \right] \right. \nn\\
&\left. \times \left(\frac{1}{\epsilon}+\ln{\frac{\mu^2 b_T^2}{b_0^2}}\right)  +\frac{4}{\epsilon^2} +\frac{2}{\epsilon}\ln{\frac{\mu^2 b_T^2}{b_0^2}} -\frac{\pi^2}{3}  \right\}.
\end{align}
The global soft function for the jet-jet dipole has already been derived in ref.~\cite{delCastillo:2020omr} and is free from rapidity divergences, yielding
\begin{align}
    \omega_{12}^{\text{global}}(\vec{b}_T, \eta_1, \eta_2, \epsilon)
    &= \frac{\alpha_s(\mu) \mu^{2 \epsilon} e^{\epsilon\gamma_E} B^\epsilon}{\pi} \Gamma(-\epsilon)\biggl[\Gamma(-\epsilon) \Gamma(1+\epsilon)\left(\frac{1+A_b}{-A_b}\right)^\epsilon \nn\\
     &\quad- \frac{\Gamma(-1-\epsilon)}{\Gamma(-\epsilon)} A_b \,{}_2F_1\left(1,1,2+\epsilon,-A_b\right)\biggr] \nn\\
    &= \frac{\alpha_s(\mu)}{4\pi} \biggl\{\frac{4}{\epsilon ^2} + \frac{4}{\epsilon} \left[\ln\frac{\mu^2b_0^2}{b_T^2}-\ln(-A_b)\right] + 2\ln^2\frac{\mu^2b_0^2}{b_T^2} -4 \ln(-A_b)\ln\frac{\mu^2b_0^2}{b_T^2} \nn\\
    &\qquad + 2 \ln^2(-A_b) -4 \ln(-A_b) \ln(1+A_b) -4\text{Li}_2(-A_b)  +\pi ^2 \biggr\},
\end{align}
with the shorthand notation
 \begin{equation}
     A_b\equiv\frac{b_T^2\left(n_1 \cdot n_2\right)}{2(n_1 \cdot b)(n_2 \cdot b)}=-\frac{\hat{s}}{4 p_T^2 \cos^2\phi_b} = -\frac{1+\cosh{(\eta_1-\eta_2)}}{2\cos^2\phi_b}, 
     \quad B\equiv\frac{|\vec{b}_T|^2}{4}.
 \end{equation}

\section{Collinear-soft and ultra-collinear-soft functions up to NNLO} \label{app:cs_ucs}
In this appendix, we present the explicit bare expression for $S_i$ defined in eq.~\eqref{eq:soft_refact1}, which combines the collinear-soft and ultra-collinear-soft contributions, up to NNLO accuracy.

Following eqs.~\eqref{eq:cs_def} and \eqref{eq:ucs_def}, the $b$-parameter dependence of $S_i$ can be reorganized as
\begin{align}
    S_i(b_\perp,\,b^+) &= 1 + \frac{Z_\alpha\alpha_s(\mu)}{4\pi} S_i^{(1)}(b_\perp, b^+) + \left( \frac{Z_\alpha\alpha_s(\mu)}{4\pi} \right)^{\!2} S_i^{(2)}(b_\perp,\,b^+)  + \mathcal{O}(\alpha_s^3),
\end{align}
where $b_\perp = b_y = b_T \sin{\phi_b}$, and 
\begin{align*}
    b^+ = b\cdot n_i = 
    \begin{cases}
        -b_T\cos{\phi_b}\sech{\eta_1}, & i=1 \\
        \,b_T\cos{\phi_b}\sech{\eta_2}, & i=2
    \end{cases}\quad.
\end{align*}
For brevity, we suppress several nonessential arguments in the notation for $S_i$ and for the collinear-soft and ultra-collinear-soft functions introduced below, including the regulators $\epsilon$ and $\eta$, the jet radius $R$, and the dependence on $\{n_i, \bar{n}_i, \underline{u}\}$.

We begin with the NLO expression,
\begin{align} \label{eq:NLO_Si}
    S_i^{(1)} (b_\perp,\,b^+) &= 
     \braket{\bm{S}^{\mathrm{ucs}(0)}_{0}\bm{S}^{\mathrm{cs}(1)}_{0}} +
    \braket{\bm{S}^{\mathrm{ucs}(1)}_{0} \bm{S}^{\mathrm{cs}(0)}_{0}} +
    \braket{\bm{S}^{\mathrm{ucs}(0)}_{1}\otimes\bm{S}^{\mathrm{cs}(1)}_{1}} \\
    &= \langle \bm{S}^{\mathrm{ucs}(1)}_{0}  \rangle + \langle \bm{1} \otimes \bm{S}^{\mathrm{cs}(1)}_{1} \rangle, \nn
\end{align}
where the superscripts $(0)$ and $(1)$ denote the perturbative order of the collinear-soft and 
ultra-collinear-soft functions. Note that the \emph{sub}scripts 0 and 1 here denote the relevant values for $m_i$ contributing at NLO and are unrelated to the perturbative order or the index $i$. The dependence on the index $i$ is hidden, on the RHS, in the suppressed arguments for the collinear-soft and ultra-collinear-soft functions.
Only the tree-level contribution of $\bm{S}^{\mathrm{ucs}}_{1}$ is needed at NLO, as shown in 
eq.~\eqref{eq:NLO_Si}, because $\bm{S}^{\mathrm{cs}}_{m_i} \sim \mathcal{O}(\alpha_s^{m_i})$.  
The second equality in eq.~\eqref{eq:NLO_Si} follows from 
$\bm{S}^{\mathrm{ucs}}_{1} = \bm{1} + \mathcal{O}(\alpha_s)$ and $\bm{S}^{\mathrm{cs}}_{0} = \bm{1}$.  
Since no additional collinear parton directions are integrated over, we omit the 
$\otimes\,\bm{S}^{\mathrm{cs}}_{0}$ term inside the color trace.

The explicit NLO expressions for the collinear-soft and ultra-collinear-soft functions are
\begin{align} \label{eq:NLO_Si2}
    \langle \bm{1} \otimes \bm{S}^{\mathrm{cs}(1)}_{1}(b_\perp) \rangle &= C_i \left( \frac{\mu |b_\perp|}{b_0}\right)^{\!2\epsilon} \left( \frac{\nu |b_\perp|R_i}{b_0}\right)^{\!\eta} h_{\rm in}, \quad
    \langle \bm{S}^{\mathrm{ucs}(1)}_{0}(b^+) \rangle = C_i \left( \frac{i\, b^+ \mu}{b_0 R_i}\right)^{\!2\epsilon}  s_{\rm out},
\end{align}
where $b_0=2e^{-\gamma_E}$, $R_i=R/(2\cosh{\eta_i})$, and 
\begin{align}
    h_{\rm in} &= \frac{2}{\epsilon^2} - \frac{4}{\eta\epsilon} - \frac{\pi^2}{6} 
    +\left(-\frac{\eta}{\epsilon^3} + \frac{\pi^2\eta}{12\epsilon} - \frac{\zeta_3}{3}\eta- \frac{\pi^2\epsilon}{3\eta} - \frac{4\zeta_3}{3}\epsilon - \frac{17\pi^4}{1440}\eta\epsilon \right. \notag\\
    &\quad\left.- \frac{4\zeta_3\epsilon^2}{3\eta} - \frac{3\pi^4}{80}\epsilon^2 - \frac{\pi^4\epsilon^3}{40\eta} \right) , \\
    s_{\rm out} &= -\frac{2}{\epsilon^2} - \frac{\pi^2}{2} +\left(- \frac{14\zeta_3}{3}\epsilon - \frac{7\pi^4}{48} \epsilon^2\right).
\end{align}

Applying similar simplifications as in eq.~\eqref{eq:NLO_Si}, one can obtain the NNLO expression of the combination of collinear-soft and ultra-collinear-soft functions,  
\begin{align}
    S_i^{(2)}(b_\perp,\,b^+)  &= \braket{\bm{1}\otimes\bm{S}^{\mathrm{cs}(2)}_{1}} + \braket{\bm{1}\otimes\bm{S}^{\mathrm{cs}(2)}_{2}} + \braket{\bm{S}^{\mathrm{ucs}(2)}_{0}} + \braket{\bm{S}^{\mathrm{ucs}(1)}_{1} \otimes \bm{S}^{\mathrm{cs}(1)}_{1}}.
\end{align}
The first two terms arise purely from the collinear-soft sector. They consist of a real–virtual contribution $\braket{\bm{1}\otimes\bm{S}^{\mathrm{cs}(2)}_{1}}$ and a double-real emission contribution $\braket{\bm{1}\otimes\bm{S}^{\mathrm{cs}(2)}_{2}}$, with all the collinear-soft real emissions constrained to lie inside the jet cone,
\begin{align}
    \braket{\bm{1}\otimes\bm{S}^{\mathrm{cs}(2)}_{1}(b_\perp)} &= \left( \frac{\mu |b_\perp|}{b_0}\right)^{\!4\epsilon} \left( \frac{\nu |b_\perp|R_i}{b_0}\right)^{\!\eta} \omega^2 C_i C_A v_A^{\rm in} \,, \\
    \braket{\bm{1}\otimes\bm{S}^{\mathrm{cs}(2)}_{2}(b_\perp)} &= \left( \frac{\mu |b_\perp|}{b_0}\right)^{\!4\epsilon} \left( \frac{\nu |b_\perp|R_i}{b_0}\right)^{\!\eta} \omega^2 
    \biggl[\omega^2 \left( \frac{\nu |b_\perp|R_i}{b_0}\right)^{\!\eta} C_i^2 \frac{h_{\rm in}^2}{2} \nn\\
    &\quad + C_i C_A h_{A} + C_i n_f T_F h_{f}\biggr].
\end{align}
The coefficients $v_A^{\rm in}$, $h_{\rm in}^2/2$, $h_A$ and $h_f$ can be computed using the parametrization in ref.~\cite{Bell:2018oqa},
\begin{align}
     v_A^{\rm in} &= -\frac{1}{\epsilon^4} + \frac{4}{\eta \epsilon^3} + \frac{\pi^2}{2} \frac{1}{\epsilon^2} - \frac{2 \pi^2}{3} \frac{1}{\eta \epsilon} + \frac{28 \zeta_3}{3} \frac{1}{\epsilon} + \frac{8 \zeta_3}{3} \frac{1}{\eta} + \frac{19 \pi^4}{120}, \\
     \frac{h_{\rm in}^2}{2} &= \frac{6}{\epsilon^4} - \frac{8}{\eta \epsilon^3} + \left( \frac{8}{\eta^2} - \frac{\pi^2}{3} \right) \frac{1}{\epsilon^2}  + \frac{4\pi^2}{3\eta^2}  + \frac{8 \zeta_3}{3 \eta} - \frac{\pi^4}{60}, \\
     h_{A} &=  \frac{1}{\epsilon^4} - \frac{4}{ \eta  \epsilon^3} +  \frac{11}{6\epsilon^3} - \frac{22}{3 \eta  \epsilon^2}  + \left(\frac{67}{18} - \pi^2\right)\frac{1}{\epsilon^2} + \left(-\frac{134}{9} + \frac{4 \pi^2}{3}\right)\frac{1}{ \eta \epsilon}  \notag\\
    &\quad  + \left(\frac{211}{27} - \frac{77 \pi^2}{36} - \frac{61 \zeta_3}{3}\right)\frac{1}{\epsilon} + \left(-\frac{808}{27} - \frac{11 \pi^2}{9} + \frac{76 \zeta_3}{3}\right)\frac{1}{\eta} -65.260(36), \\
    h_{f} &= -\frac{2}{3 \epsilon^3} + \frac{8}{3 \eta \epsilon^2} - \frac{10}{9\epsilon^2}  + \frac{40}{9\eta \epsilon}  +  \left(-\frac{74}{27} + \frac{7 \pi^2}{9}\right)\frac{1}{\epsilon}  + \left(\frac{224}{27}+\frac{4\pi^2}{9}\right)\frac{1}{\eta} + 24.72953(77).
\end{align}

The pure ultra-collinear-soft contribution takes the form
\begin{align}
    \braket{\bm{S}^{\mathrm{ucs}(2)}_{0} (b^+)} 
    = \left( \frac{i\, b^+ \mu}{b_0 R_i}\right)^{\!4\epsilon} \left[C_i^2 \frac{s_{\rm out}^2}{2} + C_i C_A s_A + C_i n_f T_F s_f\right],
\end{align}
with 
\begin{align}
    \frac{s_{\rm out}^2}{2} =  \frac{2}{\epsilon^4} + \frac{\pi^2}{\epsilon^2} + \frac{28\zeta_3}{3\epsilon}+ \frac{5\pi^4}{12}.
\end{align}
While the collinear-soft radiation is confined within the jet cone, the ultra-collinear-soft partons are unconstrained by the jet boundary and can be emitted both inside and outside the jet. However, only the out-of-cone ultra-collinear-soft momentum is measured, which leads to a further decomposition of the coefficients $s_A$ and $s_f$ as
\begin{align} \label{eq:sep_sA_sf}
    s_A \equiv g_A + v_A^{\rm out} + q_A, \quad s_f \equiv g_f + q_f.
\end{align}
The terms $g_A$ and $g_f$ denote the double-real emission contributions where both ultra-collinear-soft partons are radiated outside the jet cone and are therefore fully measured. These evaluate to
\begin{align}
    g_A &= -\frac{1}{\epsilon^4} - \frac{11}{6 \epsilon^3} + \left(-\frac{67}{18} - \pi^2\right)\frac{1}{\epsilon^2} + \left(-\frac{193}{27} - \frac{11 \pi^2}{4} - \frac{35 \zeta_3}{3}\right)\frac{1}{\epsilon} \nn \\
    &\quad + \left( -\frac{1196}{81} - \frac{67 \pi^2}{12} - \frac{31 \pi^4}{40} - \frac{473 \zeta_3}{9} \right), \\
    g_f &= \frac{2}{3 \epsilon^3} + \frac{10}{9 \epsilon^2} + \left(\frac{38}{27} + \pi^2\right)\frac{1}{\epsilon} + \left( \frac{238}{81} + \frac{5 \pi^2}{3} + \frac{172 \zeta_3}{9} \right).
\end{align}
The term $v_A^{\rm out}$ in eq.~\eqref{eq:sep_sA_sf} corresponds to the real-virtual correction with the single ultra-collinear-soft gluon radiated outside the jet cone, yielding
\begin{align}
    v_A^{\rm out} &= \frac{1}{\epsilon^4} + \frac{5\pi^2}{6}\frac{1}{\epsilon^2} + \frac{56\zeta_3}{3} \frac{1}{\epsilon} + \frac{113\pi^4}{120}.
\end{align}
Explicit forms of $g_A$, $g_f$ and $v_A^{\rm out}$ can also be extracted from the hemisphere soft function in ref.~\cite{Kelley:2011ng}. The quantities $q_f$ and $q_A$ in eq.~\eqref{eq:sep_sA_sf} describe the double-real contributions in which one ultra-collinear-soft emission lies inside the jet cone and is unobserved, while the other is emitted outside the jet cone and measured,
\begin{align}
    q_A &= \left(-\frac{2}{3}+\frac{22\pi^2}{9} - 4\zeta_3\right)\frac{1}{\epsilon} + \frac{40}{9} -\frac{134\pi^2}{27} + \frac{44\zeta_3}{3} +\frac{8\pi^4}{45},   \\
    q_f &= \left(\frac{4}{3} - \frac{8\pi^2}{9}\right)\frac{1}{\epsilon} -\frac{68}{9} + \frac{64\pi^2}{27} -\frac{16\zeta_3}{3}. 
\end{align}

The opposite-side NNLO contribution with different modes is obtained through the convolution between the NLO collinear-soft and NLO ultra-collinear-soft functions,
\begin{align}
    \braket{\bm{S}^{\mathrm{ucs}(1)}_{1} \otimes \bm{S}^{\mathrm{cs}(1)}_{1}} = 
  \left( \frac{\mu |b_\perp|}{b_0}\right)^{\!2\epsilon} \left( \frac{i b^+\mu }{b_0 R_i}\right)^{\!2\epsilon} \left[ \left( \frac{\nu |b_\perp|R_i}{b_0}\right)^{\!\eta}\omega^2 C_i^2 p_{2i} +   C_i C_A p_A\right],
\end{align}
with
\begin{align}
    p_{2i} &= h_{\rm in} s_{\rm out} = -\frac{4}{\epsilon^4} + \frac{8}{\eta \epsilon^3} - \frac{2 \pi^2}{3 \epsilon^2}  + \frac{8 \pi^2}{3 \eta \epsilon} - \frac{20 \zeta_3}{3 \epsilon} + \frac{64 \zeta_3}{3 \eta} - \frac{2 \pi^4}{15}, \\
    p_A &= \frac{2\pi^2}{3\epsilon^2} + \frac{8\zeta_3}{\epsilon}+\frac{23\pi^4}{45}.
\end{align}

Combining these pieces, the NNLO contribution to $S_i$, defined in eq.~\eqref{eq:soft_refact1}, can be organized as
\begin{align} \label{eq:NGL_coft}
S_i^{(2)}(b_\perp,\,b^+)\ &= \biggl( \frac{\mu |b_\perp|}{b_0}\biggr)^{\!4\epsilon} \biggl( \frac{\nu |b_\perp|R_i}{b_0}\biggr)^{\!\eta} \omega^2 
    \biggl[\omega^2 \biggl( \frac{\nu |b_\perp|R_i}{b_0}\biggr)^{\!\eta} C_i^2 \frac{h_{\rm in}^2}{2} \nn\\
    &\qquad + C_i C_A (h_{A}+v_A^{\rm in}) + C_i n_f T_F h_{f}\biggr] \notag \\
&\quad+  \biggl( \frac{i b^+ \mu}{b_0 R_i}\biggr)^{\!4\epsilon} \biggl[C_i^2 \frac{s_{\rm out}^2}{2} + C_i C_A (g_A+v_A^{\rm out}) + C_i n_f T_F g_f\biggr] \notag \\
&\quad+ \biggl( \frac{i b^+\mu }{b_0 R_i}\biggr)^{\!4\epsilon} \biggl[  C_i C_A (s_A - g_A - v_A^{\rm out}) + C_i n_f T_F (s_f - g_f)\biggr] \nn\\
&\quad + \biggl( \frac{\mu |b_\perp|}{b_0}\biggr)^{\!2\epsilon} \biggl( \frac{i b^+\mu }{b_0 R_i}\biggr)^{\!2\epsilon} \biggl[ \biggl( \frac{\nu |b_\perp|R_i}{b_0}\biggr)^{\!\eta}\omega^2 C_i^2 h_{\rm in} s_{\rm out} +   C_i C_A p_A\biggr]\,.  
\end{align}

\subsection{Renormalization and anomalous dimensions in multiplicity space} \label{sec:renorm_cs_ucs}
Due to the restricted phase space in the collinear-soft function, non-global logarithms (NGLs) arise in the calculation of the collinear-soft and ultra-collinear-soft functions. As discussed in~\cite{Becher:2016mmh, Becher:2016omr}, both global and non-global large logarithms can be resummed through the RG evolution of these functions.

However, part of the infrared divergence in the collinear-soft function should be removed by the collinear-soft function with lower parton multiplicity, which means its renormalization constant is a matrix in multiplicity space,
\begin{align} \label{eq:cs_renormalization}
    \bm{\mathcal{S}}_{m}^{\rm cs}(\{n_i,\bar n_i, \underline{u}\}, b_y, R, \epsilon,\eta) = 
    \sum_{l=0}^m \bm{Z}^{\rm cs}_{ml} (\{n_i,\bar n_i, \underline{u}\}, b_y, R, \epsilon, \eta, \mu,\nu)\, \bm{\mathcal{S}}_{l}^{\rm cs}(\{n_i,\bar n_i, \underline{u}\}, b_y, R, \mu,\nu),
\end{align}
where we suppress the subscript $i$ on the multiplicity indices $m_i$ and $l_i$ for notational simplicity, a convention we maintain hereafter.
In this equation, the renormalization constant $\bm{Z}^{\rm cs}$ is a lower-triangular matrix in multiplicity space. This differs slightly from the notation used in refs.~\cite{Becher:2016mmh, Becher:2016omr}, as we organize the expression such that $\bm{Z}^{\rm cs}$ multiplies the renormalized collinear-soft function from the left, equating to the bare collinear-soft function, rather than multiplying from the right.
Moreover, although eq.~\eqref{eq:cs_renormalization} includes the renormalization of rapidity divergences, the rapidity evolution is trivial and does not induce mixing between collinear-soft functions of different multiplicities. 
Thus, we can further decompose $\bm{Z}^{\rm cs}_{ml}$ into an overall factor $Z^{\rm cs}$ and a multiplicity-mixing matrix contribution $\bm{\hat Z}_{ml}$, according to
\begin{align} \label{eq:cs_Zfact}
    \bm{Z}^{\rm cs}_{ml} (\{n_i,\bar n_i, \underline{u}\}, b_y, R, \epsilon, \eta, \mu,\nu) = Z^{\rm cs}(\{n_i,\bar n_i\}, b_y, R, \epsilon, \eta, \mu,\nu) \bm{\hat Z}_{ml} (\{n_i,\bar n_i, \underline{u}\},\epsilon, \mu).
\end{align}

Similarly, the ultra-collinear-soft function can be renormalized using a lower-triangular renormalization constant $\bm{Z}^{\rm ucs}$,
\begin{align}
    \bm{\mathcal{S}}_{l}^{\rm ucs}(\{n_i,\bar n_i, \underline{u}\}, b_x, R, \mu) = \sum_{m=l}^{\infty} \bm{\mathcal{S}}_{m}^{\rm ucs}(\{n_i,\bar n_i, \underline{u}\}, b_x, R, \epsilon) \hat{\otimes} \bm{Z}^{\rm ucs}_{ml} (\{n_i,\bar n_i, \underline{u}\}, b_x, R, \epsilon, \mu),
\end{align}
which implies that higher-multiplicity ultra-collinear-soft functions are required to absorb the divergences of the lower-multiplicity ones. The symbol $\hat{\otimes}$ indicates that an integration must be performed over the additional $(m-l)$ directions of the collinear-soft partons on which the bare function $\bm{\mathcal{S}}_{m}^{\rm ucs}(\{n_i,\bar n_i, \underline{u}\}, b_x, R, \epsilon)$ depends. The renormalization constant for the ultra-collinear-soft function can likewise be decomposed into two parts,
\begin{align} \label{eq:ucs_Zfact}
    \bm{Z}^{\rm ucs}_{ml} (\{n_i,\bar n_i, \underline{u}\}, b_x, R,, \epsilon, \mu) = \frac{\bm{\hat Z}_{ml} (\{n_i,\bar n_i, \underline{u}\}, \epsilon, \mu)}{Z^{\rm ucs}(\{n_i,\bar n_i\}, b_x, R, \epsilon, \mu)} .
\end{align}
Since the multiplicity mixing in the RG evolution occurs only between the collinear-soft and ultra-collinear-soft functions, the lower-triangular matrices $\bm{\hat Z}_{ml}$ in eqs.~\eqref{eq:cs_Zfact} and \eqref{eq:ucs_Zfact} are identical and independent from the $b$-dependence.

Next, we derive the anomalous dimensions and RG equations for the collinear-soft and ultra-collinear-soft functions. The anomalous dimensions can be obtained directly from their corresponding renormalization constants. 
If the renormalization constant is a scalar, one has
\begin{align}
    \Gamma = -\lim_{\epsilon\to 0} Z^{-1}(\epsilon, \mu) \frac{\df}{\df \ln{\mu}} Z(\epsilon, \mu),
\end{align}
where $\Gamma\in \{\Gamma^{\rm cs},\, \Gamma^{\rm ucs}\}$ and $Z\in \{Z^{\rm cs},\, Z^{\rm ucs}\}$.
If the renormalization constant is a matrix in multiplicity space, the anomalous dimensions can be obtained from the following relation
\begin{align}
    \frac{\df }{\df\ln\mu} \bm{Z}_{ml}(\epsilon, \mu) = -\sum_{k=l}^m \bm{Z}_{mk} (\epsilon, \mu)\, \bm{\Gamma}_{kl}(\mu),
\end{align}
where $\bm\Gamma_{kl}\in \{\bm\Gamma^{\rm cs}_{kl},\, \bm\Gamma^{\rm ucs}_{kl}, \bm{\hat{\Gamma}}_{kl}\}$ and $\bm{Z}_{mk}\in \{\bm{Z}^{\rm cs}_{mk},\, \bm{Z}^{\rm ucs}_{mk}, \, \bm{\hat{Z}}_{mk}\}$.

Therefore, the standard form of the RG equations can be derived as,
\begin{align} 
    \frac{\df}{\df \ln\mu} \bm{\mathcal{S}}_{m}^{\rm cs}(\{n_i,\bar n_i, \underline{u}\}, b_y, R, \mu,\nu)  
    &= \sum_{l=0}^m \bm{\Gamma}^{\rm cs}_{ml}\, \bm{\mathcal{S}}_{l}^{\rm cs}(\{n_i,\bar n_i, \underline{u}\}, b_y, R, \mu,\nu) \label{eq:separate_ad_cs} \\
    &= \sum_{l=0}^m \left(\Gamma^{\rm cs} \delta_{ml} \bm{1} + \bm{\hat{\Gamma}}_{ml}\right)\, \bm{\mathcal{S}}_{l}^{\rm cs}(\{n_i,\bar n_i, \underline{u}\}, b_y, R, \mu,\nu), \nn\\
    \frac{\df}{\df\ln\mu} \bm{\mathcal{S}}_{l}^{\rm ucs}(\{n_i,\bar n_i, \underline{u}\}, b_x, R, \mu) 
    &= -\sum_{m=l}^{\infty} \bm{\mathcal{S}}_{m}^{\rm ucs}(\{n_i,\bar n_i, \underline{u}\}, b_x, R, \mu) \hat{\otimes} \bm{\Gamma}^{\rm ucs}_{ml} \label{eq:separate_ad_ucs} \\
    &= \sum_{m=l}^{\infty} \bm{\mathcal{S}}_{m}^{\rm ucs}(\{n_i,\bar n_i, \underline{u}\}, b_x, R, \mu) \hat{\otimes} \left( \Gamma^{\rm ucs} {\delta}_{ml}\bm{1} - \bm{\hat{\Gamma}}_{ml} \right). \nn
\end{align}
The second equality follows from separating the overall and multiplicity-mixing anomalous dimensions. 
Since the multiplicity-mixing contributions in the RG evolution of the collinear-soft and ultra-collinear-soft functions cancel each other, their combination $S_i$, defined in eq.~\eqref{eq:soft_refact1}, evolves under the RG 
without any residual multiplicity mixing,
\begin{align}  
    \frac{\df}{\df\ln\mu}S_i (\vec{b}_T, \eta_i, R,\mu,\nu)  = \left( \Gamma^{\rm cs} + \Gamma^{\rm ucs}\right) S_i (\vec{b}_T, \eta_i, R,\mu,\nu).
\end{align}

Based on the explicit NNLO expressions in appendix~\ref{app:cs_ucs}, the two-loop global anomalous dimensions for the collinear-soft and ultra-collinear-soft functions are
\begin{align}
    \Gamma^{\rm cs}(\alpha_s, \{n_i,\bar n_i\}, R, \mu, \nu) &= 2C_i \gamma_{\rm cusp}(\alpha_s) \ln\frac{\mu}{\nu R_i} , \label{eq:ad_cs}\\
    \Gamma^{\rm ucs}(\alpha_s, \{n_i,\bar n_i\}, b_x, R, \mu) &= -C_i \gamma_{\rm cusp}(\alpha_s) \left[ \ln{\frac{4\mu^2 b_x^2}{b_0^2 R^2}} -i\pi \text{Sign}(b_x n_{i,x}) \right], \label{eq:ad_ucs}
\end{align}
with $L_b = \ln{(\mu^2 b_T^2/b_0^2)} = \ln{(\mu^2 b_T^2e^{2\gamma_E}/4)}$. 
The treatment of the matrix-valued anomalous dimension $\bm{\hat{\Gamma}}_{ml }$
 and the final resummation of both global and NGLs is discussed in section~\ref{sec:qT_resum}.

 $S_i(\vec{b}_T, \eta_i, R,\mu,\nu)$ also satisfies the following RRG equation,
\begin{align}
    \frac{\mathrm d}{\mathrm d \ln\nu} S_i(\vec{b}_T, \eta_i, R,\mu,\nu) &= \Gamma^{S_i}_\nu ~ S_i(\vec{b}_T, \eta_i, R,\mu,\nu)
\end{align}
where the rapidity anomalous dimension is given by $\Gamma^{S_i}_\nu = - \Gamma_\nu^i (b_y, \mu)$, with $\Gamma_\nu^i (b_y, \mu)$ defined in eq.~\eqref{eq:RRG_AD}.

\section{Jet radius power corrections from hemisphere soft functions} \label{app:hemisphere}
In addition to the effective theory framework developed in \cite{Becher:2015hka, Becher:2016mmh, Becher:2016omr}, one can compute $S_i$ in the small-$R$ limit by treating it as a hemisphere soft function that is highly boosted along the jet direction $n_i$. Note that we are not referring to the standard hemisphere soft function for jet mass but include the measurements for our case, see eqs.~\eqref{eq:measure_1} and \eqref{eq:measure_2}. This allows us to assess the power corrections to the refactorization in terms of collinear-soft and ultra-collinear-soft functions.

Since the original Wilson lines $n_i$ and $\bar{n}_i$ are back-to-back, one can rotate them into the directions $n \,(n_a)$ and $\bar{n}\,(n_b)$ to simplify the calculations. Moreover, the integral definition of the boosted hemisphere soft function is invariant under this rotation.

\subsection{Integral definition}
The explicit form of the boosted hemisphere soft functions for the QCD leg labeled by parton $i$ is given by
\begin{align} \label{eq:coft_noAvg}
    S_i^{\text{hemi}}(b_\perp,\,b^+) &= 1 + \frac{Z_\alpha\alpha_s(\mu)}{4\pi} S_i^{\text{hemi},(1)}(b_\perp,\,b^+)    
    + \left( \frac{Z_\alpha\alpha_s(\mu)}{4\pi} \right)^{\!2} S_i^{\text{hemi},(2)}(b_\perp,\,b^+)  + \mathcal{O}(\alpha_s^3).
\end{align}

The single-real emission correction splits into two components,
\begin{align}
    S_i^{\text{hemi},(1)}(b_\perp,\,b^+) = S_i^{\rm in}(b_\perp) + S_i^{\rm out}(b^+),
\end{align}
whose integral definitions are
\begin{align}
    S_i^{\rm in,\, out}(b) = 16\pi^2 \left(\frac{\mu^2 e^{\gamma_E}}{4\pi}\right)^{\!\epsilon} \int \frac{\df^{d} k}{(2\pi)^{d-1}}\, \delta(k^2) \theta(k^0)  \frac{4C_i}{k^- k^+} \mathcal{M}_1^{\rm in,\, out}(b).
\end{align}
The corresponding measurement function $\mathcal{M}_1^{\rm in,\, out}$, together with the rapidity regulator (when needed), is given by
\begin{align} \label{eq:measure_1}
    \mathcal{M}_1^{\rm in}(b_\perp)= \left(\frac{\nu}{k^-}\right)^{\!\alpha} \theta\left(R_i^2-\frac{k^+} {k^-}\right) e^{i k_y b_\perp}, \quad
    \mathcal{M}_1^{\rm out}(b^+) = \theta\left(\frac{k^+} {k^-}-R_i^2\right) e^{-i \frac{1}{2}k^- b^+}.
\end{align}
Our NNLO hemisphere soft function is therefore,
\begin{equation}
    S_{i}^{\text{hemi},(2)}(b_\perp,\,b^+) = S_{i,RR}(b_\perp,\,b^+) + S_{i,RV}^{\rm in}(b_\perp) + S_{i,RV}^{\rm out}(b^+),
\end{equation}
with the real-virtual correction given by
\begin{align}
    S_{i,RV}^{\rm in, \, out}(b) = 256\pi^4 \left(\frac{\mu^2 e^{\gamma_E}}{4\pi}\right)^{\!2\epsilon}\int \frac{\df^{d} k}{(2\pi)^{d-1}}\, \delta(k^2) \theta(k^0) |\mathcal{A}_{RV}(k)|^2 \mathcal{M}_1^{\rm in,\, out}(b),
\end{align}
involving the amplitude 
\begin{align}
    |\mathcal{A}_{RV}(k)|^2 =  - (4\pi)^\epsilon\frac{C_i C_A}{4(k^-)^{1+\epsilon} (k^+)^{1+\epsilon}}  \frac{\Gamma(-\epsilon) \cot(\pi \epsilon)}{\Gamma(-2\epsilon) \sin(\pi \epsilon)}.
\end{align}
The double-real correction is given by
\begin{align}
    S_{i,RR}(b_\perp,\,b^+) &= 256\pi^4\left(\frac{\mu^2 e^{\gamma_E}}{4\pi}\right)^{\!2\epsilon} \int \frac{\df^{d} k}{(2\pi)^{d-1}}\, \delta(k^2) \theta(k^0) \int \frac{\df^{d} q}{(2\pi)^{d-1}}\, \delta(q^2) \theta(q^0) \notag \\
    &\quad \times \,\sum_X |\mathcal{A}^{(X)}_{RR}(k,q)|^2  
    \, \mathcal{M}_2^{(X)}(k,q, b_\perp, b^+),
\end{align}
where $X = C_i^2, C_i C_A, C_i n_f T_F$ denotes the color channel, and $\mathcal{M}_2(k, q, b_\perp, b^+)$ represents the measurement function combined with the Fourier factor,
\begin{align} \label{eq:measure_2}
    \mathcal{M}_2^{(X)}(k,q, b_\perp, b^+) &= \frac{1}{2!} \left[ \left(\frac{\nu}{k^-}\right)^{\!\eta} \theta\left(R_j^2-\frac{k^+} {k^-}\right) \theta\left(\frac{q^+}{q^-}-R_j^2\right)   e^{i k_y b_\perp-i\frac{1}{2} q^- b^+ } \right. \notag \\
    &\quad + \left(\frac{\nu}{q^-}\right)^{\!\eta}\theta\left(\frac{k^+}{k^-}-R_j^2\right) \theta\left(R_j^2-\frac{q^+} {q^-}\right) e^{-i\frac{1}{2} k^- b^+ + i q_y b_\perp} \notag \\
    &\quad + R^{(X)}(\nu,\eta,k^-,q^-)\, \theta\left(R_j^2 - \frac{k^+}{k^-}\right) \theta\left(R_j^2-\frac{q^+} {q^-}\right) e^{i (k_y + q_y) b_\perp} \notag \\
    &\quad + \left.\theta\left(\frac{k^+}{k^-}-R_j^2\right) \theta\left(\frac{q^+}{q^-}-R_j^2\right) e^{-i\frac{1}{2} (k^- + q^-)b^+ }\right],
\end{align}
where $R^{(X)}(\nu,\eta,k^-,q^-)$ denotes the rapidity regulator for the case in which both double-real emissions are radiated inside the jet. To derive the correct RG equations and to consistently combine the hemisphere soft functions with the jet functions, the rapidity regulator must be introduced at the level of connected webs \cite{Chiu:2012ir}. Consequently, one needs to distinguish between the contributions from uncorrelated and correlated emissions, according to
\begin{align}
R^{(X)}(\nu,\eta,k^-,q^-)=
    \begin{cases}
       (\nu/k^-)^\eta (\nu/q^-)^\eta, & X = C_i^2 \,, \\
        [\nu/(k^-+q^-)]^\eta, & X = C_i C_A \text{ or } C_in_f T_F \,.
    \end{cases} 
\end{align}

The squared amplitudes read
\begin{align}
    |\mathcal{A}^{(C_i^2)}_{RR}(k,q)|^2 &= \frac{16\,C_i^2}{k^- k^+ q^- q^+}, \\
\left| \mathcal{A}_{RR}^{(n_f)}(k, q) \right|^2 &= 2!\times 2 C_i T_F n_f \frac{2 k \cdot q (k^+ + q^+)(k^- + q^-)-(k^+ q^- - k^- q^+)^2}{(k^+ + q^+)^2 (k^- + q^-)^2 (k \cdot q)^2},  \\
\left| \mathcal{A}_{RR}^{(C_A)}(k, q) \right|^2 &=  C_i C_A \Biggl\{ - 4\frac{k^+ (2 k^- + q^-) + q^+ (k^- + 2 q^-)}{k^+ k^- q^+ q^- (k^+ + q^+)(k^- + q^-)} + \frac{2 (1 - \epsilon) (k^- q^+ - q^- k^+)^2}{(k^+ + q^+)^2 (k^- + q^-)^2 (k \cdot q)^2}
\nonumber \\
&\; \!+\! 2\frac{(k^+)^2 q^- (2 k^- \!+\! q^-) \!+\! 2 k^+ q^+ ((k^-)^2 \!-\! k^- q^- \!+\! (q^-)^2) \!+\! k^- (q^+)^2 (k^- \!+\! 2 q^-)}{k^+ k^- q^+ q^- (k^+ \!+\! q^+)(k^- \!+\! q^-)( k \cdot q)} \Biggr\}.
\end{align}

\subsection{Boosted hemisphere soft functions}
The NLO expression of the boosted hemisphere soft function coincides with the NLO result for $S_i$ defined in eq.~\eqref{eq:NLO_Si2},
\begin{align}
    S_i^{\rm in}(b_\perp) &= C_i \left( \frac{\mu |b_\perp|}{b_0}\right)^{\!2\epsilon} \left( \frac{\nu |b_\perp|R_i}{b_0}\right)^{\!\eta} h_{\rm in}, \quad
    S_i^{\rm out}(b^+) = C_i \left( \frac{i\, b^+ \mu}{b_0 R_i}\right)^{\!2\epsilon}  s_{\rm out}.
\end{align}

The NNLO hemisphere soft function takes the following form,
\begin{align} \label{eq:hemi_coft}
S_i^{\text{hemi},(2)}(b_\perp,\,b^+)\ &= \left( \frac{\mu |b_\perp|}{b_0}\right)^{\!4\epsilon} \left( \frac{\nu |b_\perp|R_i}{b_0}\right)^{\!\eta} \omega^2 
    \biggl[\omega^2 \left( \frac{\nu |b_\perp|R_i}{b_0}\right)^{\!\eta} C_i^2 \frac{h_{\rm in}^2}{2} \nn\\
    &\quad + C_i C_A (h_{A}+v_A^{\rm in}) + C_i n_f T_F h_{f}\biggr] \notag \\
&+  \left( \frac{i\, b^+ \mu}{b_0 R_i}\right)^{\!4\epsilon} \left[C_i^2 \frac{s_{\rm out}^2}{2} + C_i C_A (g_A+v_A^{\rm out}) + C_i n_f T_F g_f\right] \notag \\
&+ \left( \frac{\mu |b_\perp|}{b_0}\right)^{\!2\epsilon} \left( \frac{i b^+\mu }{b_0 R_i}\right)^{\!2\epsilon} \left( C_iC_A r_A(r) + C_i n_f T_F r_f(r) \right) \nn\\
&+ \left( \frac{\mu |b_\perp|}{b_0}\right)^{\!2\epsilon} \left( \frac{i b^+\mu }{b_0 R_i}\right)^{\!2\epsilon}  \left( \frac{\nu |b_\perp|R_i}{b_0}\right)^{\!\eta}\omega^2 C_i^2 h_{\rm in} s_{\rm out}.
\end{align}
The third and fourth lines in eq.~\eqref{eq:hemi_coft}, which differ from the expressions in eq.~\eqref{eq:NGL_coft}, correspond to configurations in which one parton is emitted inside the jet cone while the other is emitted outside.
For the uncorrelated double-real contribution, denoted by $ C_i^2 h_{\rm in} s_{\rm out}$, the two pieces factorize according to the Non-Abelian Exponentiation (NAE) theorem.
However, for correlated emissions, the presence of two distinct scales $b_\perp$ and $b^+$ gives rise to NGLs in this configuration. Consequently, $r_A$ and $r_f$ acquire a nontrivial dependence on the ratio $r$, taking the form
\begin{align}
    r_A (r) &= \frac{2\pi^2}{3\epsilon^2} + \left(-\frac{2}{3}+\frac{22\pi^2}{9} + 4\zeta_3\right) \frac{1}{\epsilon} + C_{A,0}(r), \\ 
    r_f (r) &= \frac{1}{\epsilon} \left(\frac{4}{3} - \frac{8\pi^2}{9}\right) + C_{f,0}(r),
\end{align}
with $r$ defined as
\begin{align}
    r = \frac{b_\perp R_j}{-b^+} = \frac{R}{2}\tan{\phi_b} \,.
\end{align}

\subsection{Refactorization of the hemisphere calculation and its power corrections}
Eq.~\eqref{eq:NGL_coft} and the hemisphere result in eq.~\eqref{eq:hemi_coft} differ in their third and fourth lines, which encode the doubly correlated real-emission contribution with one emission inside the cone and the other outside. We now compare these terms after averaging over $\phi_b$. After performing the $\phi_b$ average, the third line of eq.~\eqref{eq:NGL_coft} becomes \cite{Fu:2024fgj}
\begin{align}
    \left(\frac{\mu b_T}{b_0}\right)^{\!4\epsilon} \left[R^{-2\epsilon} C_i C_A \bar{p}_A + R^{-4\epsilon}\left( C_i C_A \bar{q}_A + C_i n_f T_F \bar{q}_f \right)\right],
\end{align}
with $\bar{q}_A = q_A$, $\bar{q}_f = q_f$ and 
\begin{align}
    \bar{p}_A = \frac{2 \pi^2}{3 \epsilon^2} + \left(8 \zeta_3-\frac{4}{3} \pi^2 \ln 2\right)\frac{1}{\epsilon} + \left( \frac{13 \pi^4}{45} + \frac{4}{3} \pi^2 \ln^2 2 - 16 \zeta_3\ln 2 \right).
\end{align}

When comparing with the $\bar{r}_f$ and $\bar{r}_A$ in the $\phi_b$-averaged boosted hemisphere soft functions eq.~\eqref{eq:NGL_coft}, we find that the following identities hold up to power corrections on the jet radius $R$,
\begin{align}
    \bar{r}_A &= \bar{p}_A + R^{-2\epsilon} \bar{q}_A + \mathcal{O}(\epsilon^0 R\ln{R},\epsilon^0 R) \notag\\
    &= \frac{2 \pi^2}{3 \epsilon^2} - \frac{2 \left( 3 -11\pi^2 + 6\pi^2 \ln2  - 18 \zeta_3 \right)}{9 \epsilon} + \Big[\left( \frac{4}{3} - \frac{44 \pi^2}{9} + 8 \zeta_3 \right) \ln R  \notag\\
    &\quad + \frac{40}{9} + \frac{7 \pi^4}{15} + \frac{\pi^2}{27} \left( -134 + 36 \ln^2 2 \right) + \frac{44 \zeta_3}{3}- 16 \zeta_3 \ln2  \Big] + \mathcal{O}(\epsilon^0 R\ln{R},\epsilon^0 R) ,\\
    \bar{r}_f &= R^{-2\epsilon} \bar{q}_f + \mathcal{O}(\epsilon^0 R\ln{R},\epsilon^0 R) \\
    &= \left(\frac{4}{3} - \frac{8 \pi^2}{9}\right)\frac{1}{\epsilon} + \frac{8}{9} \left(-3 + 2 \pi^2\right) \ln R + \frac{4}{27} \left(-51 + 16 \pi^2 - 36 \zeta_3\right) + \mathcal{O}(\epsilon^0 R\ln{R},\epsilon^0 R) . \nn
\end{align}

\begin{figure}[t]
    \centering
    \begin{subfigure}[t]{0.48\textwidth}
        \centering
        \includegraphics[width=\linewidth]{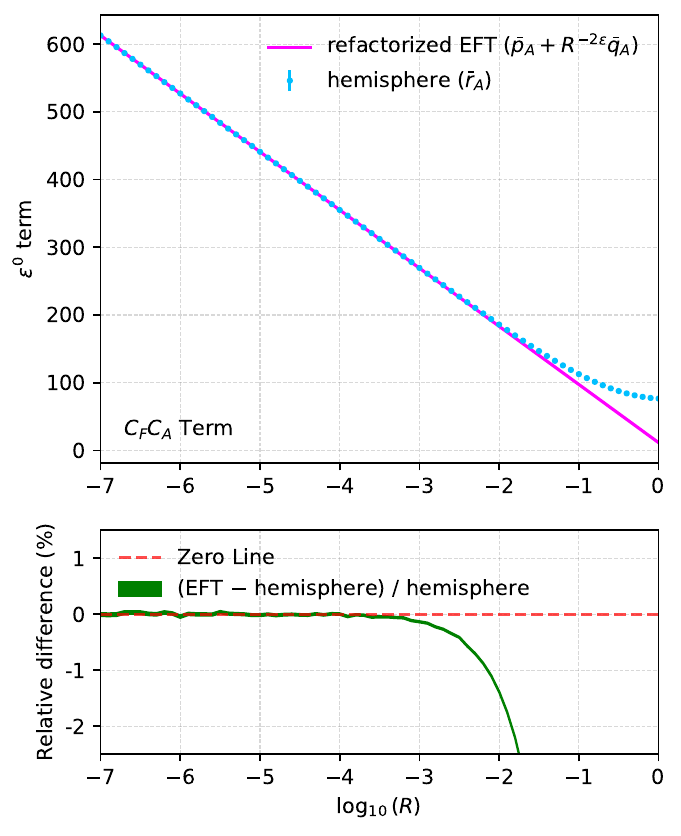}
        \caption{}
        \label{fig:CA_comp}
    \end{subfigure}
    \begin{subfigure}[t]{0.49\textwidth}
        \centering
        \includegraphics[width=\linewidth]{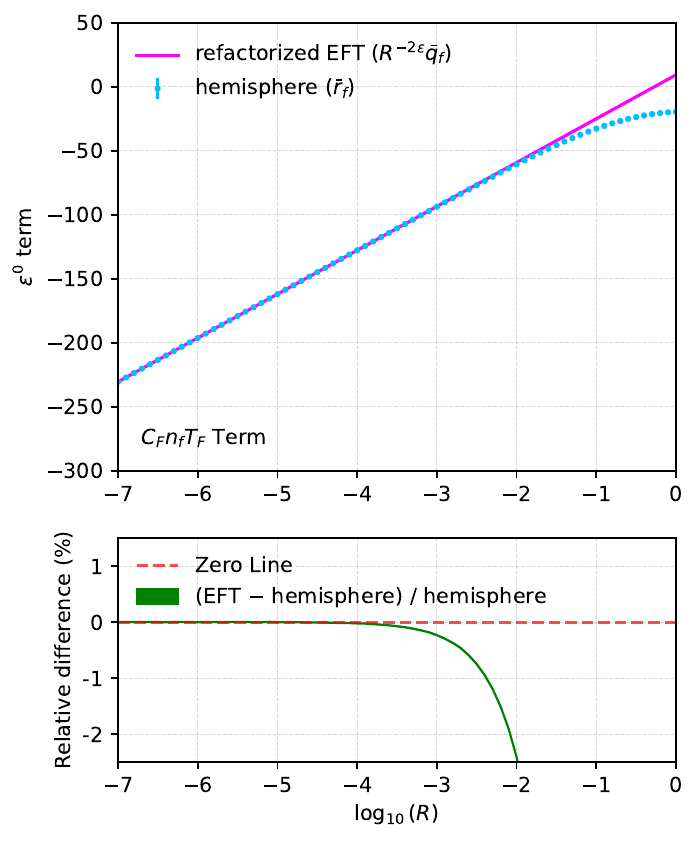}
        \caption{}
        \label{fig:nf_comp}
    \end{subfigure}
    
    \caption{Comparison of the $\mathcal{O}(\epsilon^0)$ terms of the $C_F C_A$ and $C_F n_f T_F$ color structure for the hemisphere soft function with the corresponding refactorized 
    expressions in the multi-Wilson-line EFT framework. Panel (a) compares $\bar{r}_A$ against $\bar{p}_A + R^{-2\epsilon}\bar{q}_A$, while panel (b) compares $\bar{r}_f$ against $R^{-2\epsilon}\bar{q}_f$,
    where $\bar{r}_{A,f}$ corresponds to the hemisphere soft function.}
    \label{fig:CAnf_comp}
\end{figure}

We compare the constant terms of $\bar{r}_A$ ($\bar{r}_f$) with those of $\bar{p}_A + R^{-2\epsilon}\bar{q}_A$ ($R^{-2\epsilon}\bar{q}_f$) by plotting them, as shown in figure~\ref{fig:CAnf_comp}. The comparison demonstrates that the two approaches are consistent in the small radius limit $R \ll 1$. For finite jet radius, however, the hemisphere calculation contains power corrections on $R$ relative to the multiplicity-interacting framework.

\section{Anomalous dimensions} \label{app:anomalous}
The perturbative expansions of the cusp, non-cusp, and rapidity anomalous dimensions are
\begin{align}
\gamma_{\text{cusp}} &= \sum_{n=0}^{\infty} \left(\frac{\alpha_s}{4\pi}\right)^{n+1} \gamma_n^{\rm cusp}, \quad
\gamma_{\mu}^{i} = \sum_{n=0}^{\infty} \left(\frac{\alpha_s}{4\pi}\right)^{n+1} \gamma_n^{i}, \quad 
\gamma_{\nu}^{i} = \sum_{n=0}^{\infty} \left(\frac{\alpha_s}{4\pi}\right)^{n+1} \gamma_n^{\nu}\,.
\end{align}
The cusp anomalous dimension, computed up to the three-loop order, is given by~\cite{Korchemsky:1987wg, Moch:2004pa} 
\begin{align}
\gamma^{\rm cusp}_{0} &= 4, \nn\\
\gamma^{\rm cusp}_{1} &= \left(\frac{268}{9} - \frac{4\pi^{2}}{3}\right)C_{A} - \frac{80}{9}T_{F}n_{f}, \nn\\
\gamma^{\rm cusp}_{2} &= C_{A}^{2}\left(\frac{490}{3} - \frac{536\pi^{2}}{27} + \frac{44\pi^{4}}{45} + \frac{88}{3}\zeta_{3}\right) + C_{A}T_{F}n_{f}\left(-\frac{1672}{27} + \frac{160\pi^{2}}{27} - \frac{224}{3}\zeta_{3}\right) \nonumber \\
&\quad + C_{F}T_{F}n_{f}\left(-\frac{220}{3} + 64\zeta_{3}\right) - \frac{64}{27}T_{F}^{2}n_{f}^{2}.
\end{align}
The non-cusp anomalous dimensions for the hard function are derived from the massless quark and gluon form factors, which are known up to three-loop order \cite{Idilbi:2006dg, Becher:2006mr, Becher:2009qa, Moch:2005tm}. Explicitly, we list the results up to two loops necessary for NNLL resummation,
\begin{align}
\gamma_0^q &= -3 C_F \,, \nonumber \\
\gamma_1^q &= C_F^2 \left( -\frac{3}{2} + 2\pi^2 - 24\zeta_3 \right) + C_F C_A \left( -\frac{961}{54} - \frac{11\pi^2}{6} + 26\zeta_3 \right) + C_F T_F n_f \left( \frac{130}{27} + \frac{2\pi^2}{3} \right) \,, \nonumber \\
\gamma_0^g &= -\beta_0 = -\frac{11}{3} C_A + \frac{4}{3} T_F n_f \,, \nonumber \\
\gamma_1^g &= C_A^2 \left( -\frac{692}{27} + \frac{11\pi^2}{18} + 2\zeta_3 \right) + C_A T_F n_f \left( \frac{256}{27} - \frac{2\pi^2}{9} \right) + 4 C_F T_F n_f \,,
\end{align}
The non-cusp anomalous dimensions for the beam functions and the WTA jet function are~\cite{Lubbert:2016rku}
\begin{align}
\gamma_{0}^{B_q} = \gamma_{0}^{J_q} &= 6C_F, \nn\\
\gamma_{1}^{b_q} = \gamma_{1}^{J_q} &= C_F^2 \left(3 - 4\pi^2 + 48\zeta_3\right) + C_FC_A \left(\frac{17}{3} + \frac{44\pi^2}{9} - 24\zeta_3\right) \nn\\
&\quad+ C_FT_Fn_f \left(-\frac{4}{3} - \frac{16\pi^2}{9}\right), \nn\\
\gamma_{0}^{B_g} = \gamma_{0}^{J_g} &= 2\beta_0, \nn\\
\gamma_{1}^{B_g} = \gamma_{1}^{J_g} &= C_A^2 \left(\frac{64}{3} + 24\zeta_3\right) - \frac{32}{3}C_AT_Fn_f - 8C_FT_Fn_f.
\end{align}
The non-cusp anomalous dimensions for the soft function read~\cite{Li:2014afw}
\begin{equation}
    \gamma_{0}^{S} = 0, \quad \gamma_{1}^{S} = C_{A}\left(\frac{64}{9} - 28\zeta_{3}\right) + \beta_{0}\left(\frac{56}{9} - \frac{\pi^{2}}{3}\right).
\end{equation}
Finally, we present the non-cusp rapidity anomalous dimensions~\cite{Gehrmann:2012ze,Gehrmann:2014yya,Lubbert:2016rku,Echevarria:2015byo},
\begin{align}
    \gamma_0^\nu = 0,\quad
    \gamma_1^\nu = -C_A \left(\frac{128}{9} - 56 \zeta_3\right) - \beta_0 \frac{112}{9}.
\end{align}

\bibliographystyle{JHEP}
\bibliography{dijet.bib}

\end{document}